\documentclass{article}
\usepackage[utf8]{inputenc}
\usepackage{graphicx}
\usepackage{xcolor}
\usepackage{array}
\usepackage{url}
\usepackage{lipsum}
\usepackage{listings}
\usepackage{multirow}
\usepackage{xparse}
\usepackage{array}
\usepackage{float}
\usepackage{lineno}
\usepackage{authblk}
\usepackage{hyperref}
\usepackage[normalem]{ulem}
\usepackage[a4paper]{geometry}
\usepackage{multirow}
\usepackage{tablefootnote}
\NewDocumentCommand{\codeword}{v}{%
\texttt{{#1}}%
}
\providecommand{\tabularnewline}{\\}
\floatstyle{ruled}
\newfloat{algorithm}{tbp}{loa}
\floatname{algorithm}{\protect\algorithmname}

\newcommand{\cl}{CutLang }
\definecolor{darkgreen}{rgb}{0.2,0.6,0.1}

\def\blue{\textcolor{black}}
\def\green{\textcolor{black}}

\makeatother

\begin{document}


\title{\cl v2: Advances in a runtime-interpreted analysis description language for HEP data}
\author[1]{G. Unel}
\author[2]{S. Sekmen}
\author[3]{A. M. Toon}
\author[4]{B. Gokturk}
\author[4]{B. Orgen}
\author[5]{A. Paul}
\author[6]{N. Ravel}
\author[7]{J. Setpal}
\affil[1]{University of California at Irvine, Department of Physics and Astronomy, Irvine, USA}
\affil[2]{Kyungpook National University, Department of Physics, Daegu, South Korea}
\affil[3]{Saint Joseph University of Beirut, Dept. of Computer Software Engineering, Beirut, Lebanon}
\affil[4]{Bogazici University, Department of Physics, Istanbul, Turkey}
\affil[5]{The Abdus Salam International Centre for Theoretical Physics, Trieste, Italy}
\affil[6]{University of Ankatso, Department of Physics, Antananarivo, Madagascar }
\affil[7]{R.N. Podar School, Mumbai, India}

\date{\today }
\maketitle
\begin{abstract}
We will present the latest developments in \cl, the runtime interpreter of a recently-developed analysis description language (ADL) for collider data analysis. ADL is a domain-specific, declarative language that describes the contents of an analysis in a standard and unambiguous way, independent of any computing framework. In ADL, analyses are written in human-readable plain text files, separating object, variable and event selection definitions in blocks, with a syntax that includes mathematical and logical operations, comparison and optimisation operators, reducers, four-vector algebra and commonly used functions. Adopting ADLs would bring numerous benefits to the LHC experimental and phenomenological communities, ranging from analysis preservation beyond the lifetimes of experiments or analysis software to facilitating the abstraction, design, visualization, validation, combination, reproduction, interpretation and overall communication of the analysis contents. Since their initial release, ADL and \cl have been used for implementing and running numerous LHC analyses.  In this process, the original syntax from \cl v1 has been modified for better ADL compatibility, and the interpreter has been adapted to work with that syntax, resulting in the current release v2.  Furthermore, \cl has been enhanced to handle object combinatorics, to include tables and weights, to save events at any analysis stage, to benefit from multi-core/multi-CPU hardware among other improvements. In this contribution, these and other enhancements are discussed in details. In addition, real life examples from LHC analyses are presented together with a user manual. 
\end{abstract}

\tableofcontents

\section{Introduction: DSLs for HEP analysis}

High energy physics (HEP) collider data analyses nowadays are performed using complex software frameworks that integrate a diverse set of operations from data access to event selection, from histogramming to statistical analysis.  Mastering these frameworks requires a high level knowledge of general purpose languages and software architecture. Such requirements erect a barrier between data and the physicist who may simply wish to try an analysis idea.
\blue{Moreover, even for experienced physicists, obtaining a complete view of an analysis is difficult because the physics content (e.g. object definitions, event selections, background estimation methods, etc.) is often scattered throughout the different components of the framework. This makes developing, understanding, communicating and interpreting analyses very challenging. At the LHC, almost all analysis teams have their own frameworks. There are also frameworks like CheckMate~\cite{CheckMATE,KIM2015535,2016chep.confE.120T} and MadAnalysis~\cite{MadAnalysis,Conte_2014} for phenomenology studies, and Rivet~\cite{Waugh:2006ip,BUCKLEY20132803} focused on preserving LHC analyses with unfolded results for comparison with Monte Carlo event generator predictions.
Yet, working with multiple frameworks is an extra challenge, since each framework has a different way of implementing the physics content.}

\green{It is therefore crucial to invest time in alternative approaches aiming towards the rather elusive point of easy to learn, expressive, extensible, and effective analysis ecosystem that would allow to shift the focus away from programming technicalities to physics analysis design. One way to achieve this is via a well-constructed set of libraries in a GPL supplemented with a well-designed interfaces that intrinsically imply a standard and user-friendly analysis structure. A most promising example in this area is the Scientific Python ecosystem SciPy~\cite{scipy} which brings together a popular GPL and  a rich collection of already existing bricks of classic numerical methods, plotting and data processing tools. Frameworks can be built based on the SciPy ecosystem for effective analysis, such as Coffea framework~\cite{coffea} that provides a user interface for columnar analysis of HEP data.
}


\blue{The approach that we propose in this paper to address these difficulties is the consideration of a domain specific language (DSL) capable of describing the analysis flow in a standard and unambiguous way. A DSL could be based on a completely original syntax, or it could be based on the syntax of a general purpose language, such as Python.  The important aspect would be to provide a unique and organized way of expressing the analysis content.  Applying the DSL concept to HEP analysis was first thoroughly explored as a community initiative  by a group of experimentalists and phenomenologists in the 2015 Les Houches PhysTeV workshop led to the initial design of {\tt LHADA} (Les Houches Analysis Description Accord), to systematically document and run the contents of LHC physics analyses~\cite{Brooijmans:2016vro, Brooijmans:2018xbu, Brooijmans:2020yij}. At the same time, some of the LHADA designers were already developing \cl~\cite{Sekmen:2018ehb, Unel:2019reo}, an interpreted language directly executable on events.  Being based on the same principles, in 2019, LHADA and \cl were merged by combining the best ideas from both into a unified DSL called ``Analysis Description Language (ADL)"~\cite{adlweb}, which is described in this paper.}

\blue{While the prototyping of LHADA, \cl  and ADL was in progress, parallel efforts arose in the LHC community with the aim to improve and systematize analysis development infrastructures.  One approach views each event as a database that can be queried using a language inspired by SQL, and has been  prototyped in {\tt LINQtoROOT}~\cite{LINQtoROOT} and   {\tt FemtoCode}~\cite{femtocode}. The SQL-like model is being further explored in {\tt hep\_tables} and {\tt dataframe\_expressions}~\cite{heptables} that work together to allow easy columnar-like access to hierarchical data, and in the recent experimental language {\tt PartiQL}\,\cite{partiql} designed to inject new ideas into DSL development and its extension {\tt AwkwardQL}~\cite{AwkwardQL}, designed to perform set operations on data expressed as awkward arrays.  Another study explored building a DSL embedded within YAML to describe and manage analysis content such as definitions, event selection, histogramming as well as perform data processing. The YAML-based language was integrated into the generic Python framework {\tt F.A.S.T.}\,\cite{FAST}.}  

\blue{The focused DSL developments for analyses are relatively new, but a DSL has been long embedded within the ROOT framework~\cite{BRUN199781} under the guise of {\tt TTreeFormula}, {\tt TTree::Draw} and {\tt TTree::Scan}, which allow visual or textual representation of {\tt TTree} contents for simple and quick exploratory analysis This DSL is however limited only to simple arithmetic operations, mathematical functions and basic selection criteria.  Recently, ROOT developers introduced {\tt RDataFrame}, a tool to process and analyze columnar datasets as a modern alternative for data analysis~\cite{Piparo:2699587}. Although {\tt RDataFrame} is not a DSL itself, it implements declarative analysis by using keywords for transformations (e.g. filtering data, defining new variables) and actions (e.g. creating histograms), and is interfaced to the ROOT classes {\tt TTreeReader} and {\tt TTreeDraw}.  {\tt RDataFrame} recently led to the development of the preliminary version of another DSL and its interpreter called {\tt NAIL} (Natural Analysis Implementation Language)~\cite{NAIL}.  {\tt NAIL}, written in Python.  It takes CMS NanoAOD~\cite{nanoaod} as an input event format and generates {\tt RDataFrame}-based C++ code, either as a C++ program or as a C++ library loadable with ROOT. }

\blue{All these different approaches and developments were discussed among experimentalists, phenomenologists and computer scientists in the first dedicated workshop ``Analysis Description Languages for the LHC'' at Fermilab, in May 2019~\cite{lpcadlworkshop}. The workshop resulted in an overall agreement on the potential usefulness of DSLs for HEP analysis, elements of a DSL scope. and an inclination to pursue multiple alternatives with the ultimate goal of a common DSL for the LHC that combines the best elements of the different approaches~\cite{Sekmen:2020vph}. The activities in DSL development are therefore ongoing with a fast pace.}

\blue{This initial positive feedback has motivated further progress in ADL, which will be described here. } ADL is a declarative language that can express the mathematical and logical algorithm of a physics analysis in a human-readable and standalone way, independent of any computing frameworks.  Being declarative, ADL expresses the analysis logic without explicitly coding the control flow, and is designed to describe what needs to be done, but not how to do it.  This consequently leads to a more tidy and efficient expression and eliminates programming errors.  At its current state, ADL is capable of describing many standard operations in LHC analyses.  However, it is being continuously improved and generalized to address an even wider range of analysis operations.  

ADL is designed as a language that can be executed on data and used in real life data analyses.  An analysis written with ADL could be executed by any computing framework that is capable of parsing and interpreting ADL, hence satisfying the framework independence.  Currently, two approaches have been studied to realize this purpose.  One is the transpiler approach, where ADL is first converted into a general purpose language, which is in turn compiled into code executable on events.  A transpiler called {\tt adl2tnm} converting ADL to C++ code is currently under development~\cite{Brooijmans:2018xbu}.  Earlier prototype transpilers converting {\tt LHADA} into code snippets that could be integrated within CheckMate~\,\cite{CheckMATE,KIM2015535,2016chep.confE.120T} and Rivet~\cite{Waugh:2006ip,BUCKLEY20132803} frameworks were also studied. The other approach is that of runtime interpretation.  Here ADL is directly executed on events without being intermediately converted into a code requiring compilation.  This approach was used for developing \cl~\cite{Sekmen:2018ehb, Unel:2019reo}.  

In this paper, we focus on \cl and present in detail its current state denoted as CutLang v2, which was achieved after many improvements on the early prototype \cl v1 introduced in~\cite{Sekmen:2018ehb}. Hereafter, \cl v2 will be referred to as \cl for brevity. The main text emphasizes the novelties that led to ADL and improved \cl.  We start with an overview of ADL in Section~\ref{sec:adlfile}, then proceed with describing technicalities of runtime interpretation with \cl in Section~\ref{sec:clint}.  We next present the ADL file structure and analysis components that can be expressed by ADL, focusing on the new developments and recently added functionalities in Section~\ref{sec:anlcont}. This is followed by Section~\ref{sec:output} describing analysis output, again focusing on new additions, Section~\ref{sec:multithread}, explaining the newly-added multi-threaded run functionality, Section~\ref{sec:maintenance} on \cl code maintenance and recently incorporated continuous integration, Section~\ref{sec:AnaEx} detailing studies on analyses implementation, and conclusions in Section~\ref{sec:conclusions}.  The full description of the current language syntax is given in the form of a user manual in Appendix~\ref{appendix:A}, followed by a note on the \cl framework and external user functions in Appendix~\ref{appendix:B}.

\section{ADL overview: File and functions}
\label{sec:adlfile}


In ADL, the description of the analysis flow is done in a plain, easy-to-read text file, using syntax rules that include standard mathematical and logical operations and 4-vector algebra.  In this ADL file, object, variable, event selection definitions are clearly separated into blocks with a keyword value/expression structure, where keywords specify analysis concepts and operations. Syntax includes mathematical and logical operations, comparison and optimization operators, reducers, 4-vector algebra and HEP-specific functions (e.g. $d\phi$, $dR$).  However, an analysis may contain variables with complex algorithms non-trivial to express with the ADL syntax (e.g. $M_{T2}$~\cite{Barr:2003rg}, aplanarity) or non-analytic variables (e.g. efficiency tables, machine learning discriminators). Such variables are encapsulated in self-contained, standalone functions which accompany the ADL file.  Variables defined by these functions are referred to from within the ADL file.  As a generic rule, all keywords, operators and  function names are case-insensitive.
n
The language content, syntax rules, and working examples of self-contained functions will be presented in the coming sections, after a technical introduction of the \cl interpreter.  

\section{Technical background of the \cl interpreter}
\label{sec:clint}

An interpreted analysis system makes adding new event selection criteria, changing the execution order or cancelling analysis steps more practical.  Therefore \cl was designed to function as a runtime interpreter and bypass the inherent inefficiency of the modify-compile-run cycle.  Avoiding the integration of the analysis description in the framework code also brings the huge advantage of being able to run many alternative analysis ideas in parallel, without having to make any code changes, hence making the analysis design phase more flexible compared to the conventional compiled framework approach.  

\begin{figure}[h!]
    \centering
    \includegraphics[scale=0.3]{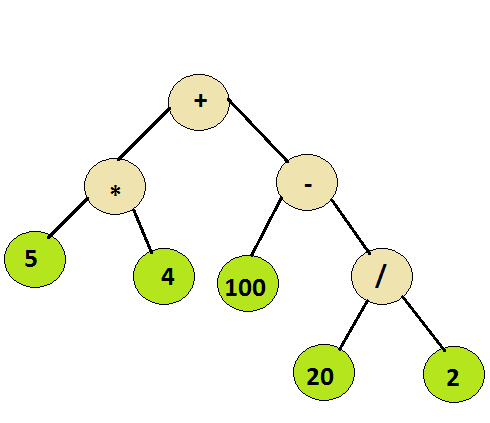}
    \caption{An expression tree example: the program traverses the tree from right to left evaluating the encountered functions from bottom to top.    \label{exptree}}
    \label{fig:my_label}
\end{figure}

\cl runtime interpreter is written in C++, around function pointer trees representing different operations such as event selection or histogramming. Therefore processing an event with a cutflow table becomes equivalent to traversing multiple expression trees with arbitrary complexities, such as the one shown in Figure~\ref{exptree}. Here physics objects are given as arguments.

Handling of the Lorentz vector operations, pseudo-random number generation, input-output file and histogram manipulations are all based on classes of the ROOT data analysis framework~\cite{ROOT}. The actual parsing of the ADL text relies on automatically generated dictionaries and grammar based on traditional Unix tools, namely, Lex and Yacc~\cite{lexyacc}. The ADL file is split into tokens by Lex, and the hierarchical structure of the algorithm is found by Yacc.  Consequently, \cl can be compiled and operated in any modern Unix-like environment. The interpreter should be compiled only once, during the installation or when optional external functions for complex variables are added. Once the work environment is set up, the remainder is mostly a think-edit-run-observe cycle. 
\blue{The parsing tools also address the issue of possible user mistakes with respect to the syntax. \cl output clearly indicates the problem, and the line number of the offending syntax error. 
However the logical inconsistencies, such as imposing a selection on the third jet's momentum while only requesting at least two jets are not yet handled. Ensuring the consistency of the algorithm needs to be done by the user.  Input and output to \cl is via ROOT files.  The description of the input files and event formats are given below while the description of the output file and its contents are given in section~\ref{sec:output}.}

\subsection{Event input}

\blue{The \cl framework takes the input event information in the ROOT ntuple format and can work with different input event data types each implemented as a plug-in. Widely used event formats such as ATLAS and CMS open data~\cite{opendata}, CMS NanoAOD~\cite{nanoaod}, Delphes~\cite{DELPHES} and LHCO event are by default recognized and can be directly used. New or custom input event formats can also be easily added via usage of event class headers via a well-defined procedure  described in Appendix~\ref{app:newformat}.  The potential changes in the existing event formats and addition of new event formats currently need to be adapted manually following the mentioned procedure. \cl has its own internal event format called {\tt LVL0}.  The contents of the input event formats including all particle types and event properties are worked through an internal abstraction layer and adapted to {\tt LVL0}, which, in turn connects to the syntax of ADL.  The purpose of this approach is to have ADL be independent of the input file format and be capable of running the same ADL analysis with \cl on any input file.  This implies that only a subset of event content is readily recognized via \cl when expressed within the ADL syntax.  However, any event variables or attributes included in the existing event files and formats can be easily added through external user functions.  This way, they can be referred to within the ADL files and be recognized by \cl.  The practical details of this procedure can be found in Appendix~\ref{app:userfunc}.}


\section{Description of the analysis contents}
\label{sec:anlcont}

We will now explain in detail which analysis components and physics algorithms can be described by ADL and processes with \cl.  We will prioritize  highlighting the many novelties added and improvements that took place since the original versions \cl v1 and LHADA.  The descriptions here concentrates on the concepts and content that can be expressed and processed by ADL and functionalities of \cl v2, rather than attempting to give a full layout of syntax rules, which is independently provided in the user manual in Appendix~\ref{appendix:A}.

\subsection{ADL file structure in \cl}
\label{sec:adlfilestr}

As a runtime interpreter, \cl processes events in a well-defined order.  It executes the commands in the ADL file from top to bottom.  Therefore, the ADL files are required to describe the analysis flow in a certain order.  Some dedicated execution commands are also used within the ADL file, in order to facilitate the runtime interpretation. Throughout the ADL file, the mass, energy and momentum are all written in Giga Electron Volt (GEV) and angles in radians. User comments and explanations should be preceded by a hash (\#) sign. 
To be executable with \cl, an ADL file would consist of five possible sections described below, out of which, existence of one section is mandatory:

\begin{description}
\item [{initializations:}] This section contains commands that are related to analysis initialization and set up, for which, the relevant keywords are summarized in Table~\ref{tab:initializations}.  The keywords and values are separated by an equal sign. The last two lines in the table refer to the lepton (electron or muon) triggers. Their utilization is described in Appendix~\ref{app:predefobj}, it is worth noting at this point that Monte Carlo (MC) simulation weights are not taken into account when the trigger value is set to data.
\item [{countformats:}] This section is used for setting up the recording of already existing event counts and errors, e.g., from an experimental paper publication. It is therefore not directly relevant for event processing, but rather for studying the interplay between the results of the current analysis and its published experimental counterpart. More generally, it is used to express any set of pre-existing counts of various signals, backgrounds, and data (together with their error) of an analysis.
\item [{definitions1:}] 
This section is used for defining aliases for objects and variables, in order to render them more easily referable and readable in the rest of the analysis description.  
For example, it can introduce shortcuts like \texttt{Zhreco} for a hadronically reconstructed Z boson, or values like mH, i.e., mass of a reconstructed Higgs boson. These definitions can only be based on the predefined keywords and objects. 
\item [{objects:}] This section can be used to define new objects based on predefined physics objects and shorthand notations declared in definitions1.
\item [{definitions2:}] This section is allocated for further alias or shorthand definitions. Definitions here can be based on objects in the previous section and predefined particles.
\item [{event categorization:}] This section is used for defining event selection regions and criteria in each region.  Running with CutLang requires having at least one selection region with at least one command, which may include either a selection criterion or a special
instruction to include MC weight factors or to fill histograms. 
\end{description}

We next describe the detailed contents and usage of these sections.

\begin{table}[h]
\caption{ Initialization keywords and their possible values\label{tab:initializations} }

\centering{}%
\begin{tabular}{|r|l|}
\hline 
Keyword & Explanation\tabularnewline
\hline 
\hline 
\texttt{SkipHistos} & Skip (=1) or Display (=0) the histograms in final efficiency table\tabularnewline
\hline 
\texttt{SkipEffs} & Skip (=1) or Display (=0) the final efficiency table \tabularnewline
\hline 
\texttt{TRGm} & 0=Off, 1=Data, 2=MC for muons\tabularnewline
\hline 
\texttt{TRGe} & 0=Off, 1=Data, 2=MC for electrons\tabularnewline
\hline
\texttt{RandomSeed} & random number generator seed, an integer \tabularnewline
\hline
\end{tabular}
\end{table}

\subsection{Object definitions}

Generally, the starting point in an analysis algorithm is defining and selecting the collections of objects, such as jets, $b$ jets, electrons, muons, taus, missing transverse energy, etc. that will be used in the next steps of the analysis.  Usually, the input events contain very generic and loose collections for objects, which need to be further refined for analysis needs. \cl is capable of performing a large variety of operations on objects, including deriving new objects via selection, combining objects to reconstruct new objects, accessing the daughters and constituents of objects.  Once an object is defined, it is also possible to find objects with a minimum and maximum of a given attribute within the object's collection, or sort the collection according to an attribute.

In the ADL notation, object collection definitions are clearly separated from the other analysis tasks.  Here the term object is used interchangeably with object collection.  Each object is defined within an individual {\tt object} block uniquely identified with the object's name.  These blocks, starting with the input object collection(s)'s name(s), list different types of operations afterwards.

\cl automatically retrieves all standard object collections from the input event file without the need for any explicit user statements within the ADL file.  It can read events with different formats, such as Delphes fast simulation~\cite{DELPHES} output, CMS NanoAOD~\cite{nanoaod}, ATLAS or CMS open data~\cite{opendata} ntuples and recognize the object collections in these.  One property unique to \cl is that it is designed to map input object collections to common, standard object names with a standard set of attributes, as described in Appendix~\ref{app:predefobj} and ~\ref{app:predeffunc}.  For example, {\tt AK4jets} collection in CMSNanoAOD and {\tt JET} collection in Delphes are both mapped to {\tt Jet}.  This approach allows to process the same ADL file on different input event formats, and has proven very useful in several simple practical applications.  However, we also recognize that this approach only works when different input collections have matching properties, e.g. when Delphes electrons and CMS electrons have to the same identification criteria which can be mapped to the same identification attribute, or a Delphes jet and an ATLAS jet use the same b-tagging algorithm that can be mapped to the same b-tagging attribute.  Therefore, other interpreters of ADL may choose to use input collection and attribute names as they are, in order to be more unambiguous.  Allowing to practice different approaches with advantages for different use cases, while still adhering to the principle of clarity is a significant aspect of ADL.  

The most common object operation is to take the input object collection and filter a subset by applying a set of selection criteria on object attributes.  This can be done very straightforwardly in ADL by listing each selection criterion in consecutive lines.  The objects in the input collection satisfying the criteria can be either selected or rejected using the {\tt select} or {\tt reject} keywords.  Comparison operators such as $=,\, !=\, >,\, <,\, >=,\, <=$ can directly be used for expressing the criteria.  Logical operators {\tt AND}, {\tt OR} and {\tt NOT} can be used for expressing composite or reverted criteria.  A complete list of these operators can be found in Appendix~\ref{app:comprangelog}.

It is also possible to filter an object collection based on other object collections, such as in the cases of object cleaning or matching.  For example, one can reject jets overlapping with photons, or select boosted W jets matching generator level W bosons.  Such operations involve intrinsic loops, which are readily handled by \cl.  Functions such as $\delta \phi$ or angular distance $\delta R$ can be readily used when comparing objects. Given an initial object collection, one can consecutively derive several objects.  For example {\tt jets} can be filtered to obtain {\tt cleanjets}, while {\tt cleanjets} can be further filtered to obtain {\tt verycleanjets}.  One can also use the same initial collection to define different collections such as taking {\tt muons} and imposing different criteria to obtain {\tt loosemuons} and {\tt tightmuons}.  

Another very common operation is to combine objects to reconstruct new objects, such as combining 2 leptons to form a $Z$ boson.  Sometimes, the reconstruction could be very straightforward, as in requesting to reconstruct only a single $Z$ boson per each event.  However, in other cases, one might have to reconstruct as many $Z$ bosons as possible.  In each case, reconstructed candidates might undergo filtering or selection of a single most optimal candidate among all candidates.  Combination operations are very diverse, and finding a completely generic expression for them is non-trivial.  In its v1, \cl could reconstruct an explicitly defined number of objects per event.  It could find the object satisfying given criteria by performing optimization operations.  In v2, \cl has been generalized to reconstruct any number of objects, by taking into account the combinatorics.  Selection criteria can also be imposed on both the input and reconstructed objects.  Technical information on how to perform combinations is provided in Appendix~\ref{app:comb}.

Another common situation is when objects in a collection are individually associated to other collections.  Examples include mothers or daughters of generator level particles, subjets or constituents of jets, associated tracks of leptons or jets.  As a first \cl was adapted in v2 to work with jet constituents using the syntax described in Appendix~\ref{app:constituents}.  Another example of association is daughters of generator truth level particles.  If an analysis if performed directly on generator level particles, or if a study is required on truth level particles, information such as PDGID codes or decay chain become relevant.  \cl is now capable of accessing PDGID and the decay products of a particle (referred as "daughters" in HEP), with the syntax described in Appendix~\ref{app:pdgid} and~\ref{app:daughters}.  \cl provides both the number of daughters and a modifier to refer to the daughters. 
Work is in progress for finding a generalizable syntax for object association expressions.  

Members of object collections can be directly accessed via their indices.  Being declarative, ADL syntax does not include explicit statements for looping over object collections, and \cl is capable of interpreting this implicit looping.  For example, when filtering a jet collection, one might apply a cleaning criterion which requires no electron to be in the proximity of the jet defined by a radius.  Applying this criterion requires looping over electrons, however it suffices to write the electron object's name in order for \cl to interpret implicit looping based on the context.  In other cases, it might be necessary to access only a subset of the collection, such as when imposing a selection on the $\delta \phi$ between first 3 jets with highest $p_T$ and the missing transverse momentum.  ADL and \cl were updated to allow such operations.  The Python slice notation has been adapted for expressing subset ranges in object collections, as described in Appendix~\ref{app:loopoversubset}.

Input or defined object collections are by default sorted by \cl in the order of decreasing transverse momentum $p_T$.  ADL can express sorting object collections according to any feature, in ascending or descending order, and \cl is capable of performing such sorting operations.  Moreover, so-called "reducers" can be applied for extracting values from existing object collections.  One case is the capability to extract the maximum or minimum value of a given attribute in an object collection.  For example, \cl can give the maximum $p_T$ possessed by a jet in a jet collection, or minimum value of isolation possessed by an electron in an electron collection.  Another case is the summation operation, where one can sum over the values of a given attribute over the whole collection.  The most common use case here is the summation of object $p_T$s to obtain event variables such as the hadronic transverse energy $H_T$.  Sorting and reducers are recent additions to ADL and \cl and the details on their implementation and usage are given in Appendix~\ref{app:sort},~\ref{app:maxmin},~\ref{app:sum} and in the examples referred to in Section~\ref{sec:AnaEx}. 

\subsection{Object or event variables}

An object variable is a quantity defined once per object, such as a jet's transverse momentum $p_T$ or an electron's relative isolation.  An event variable is a quantity defined once per event, such as missing transverse energy $E_T^{miss}$, number of electrons selected using the tight criteria, $p_T$ of the highest $p_T$ jet, transverse mass calculated using the highest $p_T$ lepton and $E_T^{miss}$.  Object and event variables used in object definitions or event categorization in an analysis are not always fully provided in the input event data.  These quantities therefore need to be computed during the analysis using the existing inputs.  ADL is designed to allow definition of such new variables in two ways.  Simple variables that could be described analytically using a single line formula can be expressed within the ADL file using mathematical operations.  A classic example would be that of the definition of transverse mass obtained from a visible object, such as a lepton, and the missing transverse energy. To enable writing these simple formulas, \cl is capable of parsing and processing operators such as $+,\,-,\,*,\,/,\,\hat{}\,$. \cl has also incorporated a series of internal functions to express other operations such as  abs(), sqrt(), sin(), cos(), tan() and log().  Reducer operators used for reducing collections to a single value, e.g. size(), sum(), min(), max() are also available for computing quantities.  For example, the hadronic transverse momentum $H_T$ can be computed from all jets in an event using the sum() reducer as {\tt sum(pT(jets))}.  

However, in many cases, variables are defined by complex algorithms non-trivial to express.  Examples such as angular separation $dR$, aplanarity, stransverse mass $M_{T2}$~\cite{Barr:2003rg}, razor variables~\cite{Rogan:2010kb}, etc. either cannot be easily written using the available operators or require multiple steps of calculation.  Some of these algorithms, like angular separation and razor variables were predefined as internal functions in \cl, and more, like $H_T$ and $M_{T2}$ were added recently.  A list of existing variables can be found in Appendix~\ref{app:predeffunc}. Other algorithms can be easily incorporated by the user following the recently generalized recipe in Appendix~\ref{appendix:B}.  Another class of sophisticated variables include quantities defined from numerical functions, such as object or trigger efficiencies used to compute object or event weights, provided in tables or histograms, or discriminators/efficiencies computed via machine learning models.  All these variables are incorporated by being defined in independent, self-encapsulated functions outside the ADL file and referring to them within the ADL file.  These external user functions should be seen as a natural extension of the language. The ultimate aim is to provide these functions in a well-defined and straightforwardly extendable database. 

The expressions for variables, whether they are built directly using the available mathematical operators or indirectly via internal or user functions, can be written openly in the place of usage, e.g. in the line when a selection is applied on the variable.  Alternatively, if the variable is used multiple times in an analysis, e.g. in different selection regions, it can be defined once, using the {\tt define} keyword, which allows to assign an alias name to the variable.  Currently, defining aliases using the {\tt define} keyword is only possible for event variables in \cl, but not for object variables.  In \cl, the {\tt define} expressions are uniquely placed at the end of the {\tt object} blocks and before the beginning of the event selection.

\subsection{Event categorization}

In a typical collider analysis, events are categorized based on different sets of selection criteria applied on event variables into a multitude of signal regions enhancing the presence of the signal of interest, or control or validation regions used for estimating backgrounds.  These regions can be derived from each other, and can be correlated or uncorrelated depending on the case.  ADL organizes event categorization by defining each selection region in an independent {\tt region} block~\footnote{This block was called {\tt algo} in the original \cl syntax.  Even though {\tt algo} is still valid in \cl, we generally refer to the block as {\tt region}, as the latter is a more domain specific word.} and labels each region with a unique name.  The {\tt region } blocks mainly consist of a list of selection criteria.  As in the case for objects, each criterion is stated in a line starting with a {\tt select} or a {\tt reject} keyword, which allows to select or reject the events satisfying the criterion, respectively.  Comparison operators, logical operators and ternary operator, syntax for which is described in Appendix~\ref{app:comprangelog} are used for expressing the criteria.  Another operation that can be performed within the context of event classification is $\chi^2$ optimization for reconstructed quantities, whose syntax is described in Appendix~\ref{app:chi2min}.  An example would be finding among several top quark candidates, the candidate with mass closest to the top quark mass, and using the optimal candidate's properties for further selection.

ADL and \cl allow deriving selection regions from each other, e.g. deriving multiple signal regions from a baseline selection region.  This is done by simply referring to the baseline region by name in the new region's block, and not repeating the whole selection every time. 

In many analyses, especially those targeting searches for new physics, events in given search regions are partitioned into many bins based on one or more variables, e.g. $H_T$, $E_T^{miss}$ or some invariant mass.  Data counts and background estimates in these bins constitute the result of the analysis.  With the increased data, recent LHC analyses, especially inclusive searches for new physics may contain hundreds of bins. Treating each bin as an independent search region and writing a separate block for each would be highly impractical.  As an alternative, recently, the capability of binning the events in a given region was added to ADL and \cl through the {\tt bin} keyword.  Bins in a region, by definition, are to be non-overlapping. The \cl interpreter and framework operate based on this principle, and skip an event once it is classified into a bin.  This property distinguishes bins from regions, as different regions can be overlapping, and a given event is evaluated for all regions, independent of whether it is selected or not by the preceding regions.  Bins can be described in two ways: when the binning is done using only a single variable, all bins can be defined in a single line, by specifying the variable name and the bin intervals.  When bins are defined based on multiple variables, this way of description can become ambiguous, and a more explicit description, where each bin is defined in one dedicated line can be used.  The usage and syntax of the {\tt bin} keyword is described in Section~\ref{app:bins}.  In case multiple regions would have the same binning (e.g. a signal region and several control regions from which the background is estimated), currently, the binning definitions must be separately specified in each region independently.  We are searching for a more practical way of expression which would avoid the repetition, while keeping with the human readability principle.  

\subsection{Event weights}
\label{sec:evtweights}

In an analysis, events, especially simulated events are usually weighted in order to match the real data luminosity or to correct for detector effects. \cl has been recently adapted to incorporate the capability of applying event weights.  Event weights can be applied within the {\tt region} blocks via usage of the {\tt weight} keyword as described in Appendix~\ref{app:weight}.  A particular event selected by two different regions can receive different weights.  Event weights can be either constant numbers or functions of variables.  These functions may include analytical or numerical internal or user functions.  Weights based on numerical functions, such as efficiencies (e.g. trigger efficiencies) can also be applied from tables written within the ADL file, as described in Appendix~\ref{app:tables}.  The systematic way for expressing efficiencies in tables and applying them to objects and events was incorporated recently in ADL and \cl.

\subsection{Applying efficiencies to objects and events using the hit-and-miss method}
\label{sec:applyeffhm}

As mentioned above, applying efficiencies to events and objects, such as trigger efficiencies or object reconstruction, identification and isolation efficiencies is a common part of many analyses.  Section~\ref{sec:evtweights} described how to apply the effect of event efficiencies as event weights.   There is, however, another approach, which involves emulating the effects of efficiencies.  This approach involves randomly accepting events or objects having a certain property, such that the total selected percentage reflects that of the efficiency.  For example, if the overall reconstruction and identification efficiency for an electron with $20 < p_T < 40$~GeV and $|\eta| < 2.1$ is 60\%, a given MC truth electron in that $p_T$ and $|\eta|$ range is allowed to pass the selection only with a 0.6 probability.  The decision for selection is made by sampling a uniform random number between 0 and 1, and accepting the event or object if the uniform random number is greater than the efficiency value.  Usually, the uncertainty on the efficiency is also taken into account when making the pass/fail decision.  This is called the hit-and-miss method. 

Emulating efficiencies using the hit-and-miss method is regularly used in parametrized fast simulation frameworks.  It is also becoming increasingly relevant to incorporate this functionality in the analysis step, especially for the benefit of phenomenological studies targeting interpretation or testing new analysis ideas.  These studies generally use events produced by fast simulation or even at truth level instead of real collision data events or MC events produced by full detector simulation as used in experimental analyses.  Experimental analyses use complicated object identification criteria, which cannot be implemented by fast simulation.  Moreover, it is common to see different analyses working with different identification methods for a given object (e.g. cut-based identification versus multivariate analysis-based identification for electrons), as different methods may perform better for different physics cases.  Consequently, working with different phenomenology analyses each using different identification criteria requires implementing all these criteria in the simulation step, which is highly impractical.  Therefore, it is helpful for the infrastructure handling the analysis step to have the capability to emulate using efficiencies.

Emulating efficiencies with uncertainties was recently incorporated in \cl.  The hit-and-miss method is applied via the internal function {\tt applyHM}.  In the current implementation, the efficiency values and errors versus object  properties are input via {\tt table} blocks in the ADL file.  This will be generalized to reading efficiencies from other formats, e.g. input histograms or numerical external functions in the near future.  

The {\tt applyHM} function uses a uniform distribution to decide if the central value was hit (below the value) or missed (above the value), the central value itself is recalculated in case the table contains errors. The new value is recalculated each time based on a double Gaussian function with positive and negative widths which are the errors of the associated bin in the efficiency table:
\begin{equation}
    dg(x)\equiv \sqrt{\frac{2}{\pi*\epsilon_u*\epsilon_d}}*\left[e^-\frac{(x-\mu)^2}{2*\epsilon_d^2}\times\theta(\mu) + e^-\frac{(x-\mu)^2}{2*\epsilon_u^2 } \right]
\end{equation}
where $\mu$ is the central value of the relevant bin from efficiency table, $\epsilon_u$ and $\epsilon_d$ are the errors in the same bin and finally $\theta$ is the unit step function.  The {\tt applyHM} function can both be used in the {\tt object} blocks for defining derived object collections.  It can also be used in the {\tt region} blocks to apply efficiencies on a particular object, e.g. to check whether the jet with the highest $p_T$ is a b-tagged jet or not.   Syntax for the {\tt applyHM} function can be found in Appendix~\ref{app:applyHM}.  

\subsection{Histogramming}

As described in the introduction, the main scope of ADL is the description of the physics content and algorithmic flow of an analysis.  The language content presented up to this point serves this purpose.  However further auxiliary functionalities are required for practicality while running the analysis on events.  One such functionality is histogramming. Since the start of its design, \cl has been capable of filling one-dimensional histograms of event variables.  Recently, the capability of drawing two-dimensional histograms has been added.  The syntax for histogramming can be found in Appendix~\ref{app:histo}.  Histogramming is currently only available for event variables. It will be added for object properties in the near future.  

\subsection{Alternatives vocabulary and syntax}

The main priority of the ongoing developments is to establish the principles of ADL as a language. Here, we refer to a language as a set of instructions to implement algorithms that produce various kinds of output through abstractions for defining and manipulating data structures or controlling the flow of execution.  It is however important to distinguish that a language can be expressed using alternative vocabulary or syntax.  Here, vocabulary is the words with a particular meaning in the language, such as block or keyword names, and syntax is the set of rules that defines the combinations of symbols that are considered to be a correctly structured expression of the language. 
Our experience on the way from \cl v1 and LHADA to ADL showed that there might not always be a single best syntax for expressing a given content.  Alternative syntax options may be more favorable in different use cases, due to practicality or simply due to different tastes of the users.  Recognizing this, we recently opted to host multiple syntactic alternatives in ADL and \cl for several cases.  The most obvious case is the syntax for expression of object attributes, as described in Appendix~\ref{app:predefobj}. \blue{It should be noted that these alternatives can only exist for simple, localized syntactic expressions but not for the overall content and structure of the language.} A more minor example is the name for the event classification block keyword, i.e. both {\tt region} and {\tt algo} are valid.  Another is in the expression of specifying the input object collection in an {\tt object} block, where either {\tt take} keyword, {\tt using} keyword or a colon ":" are valid.  \cl was recently updated to be able to parse and interpret different alternatives in such cases. \blue{We believe such flexibility will allow users to find the best ways to express their ideas and moreover will help \cl to grow its overall user base.}


\section{Analysis output}
\label{sec:output}

\cl as an analysis framework is designed to output information and data that would be used for further analysis.  The main output obtained after running an analysis in \cl is provided in a ROOT file.  The file, first of all, includes a copy of the ADL file content in order to document the provenance of the analysis.  It also includes histograms with all the event counts and uncertainties obtained from the analysis and all histograms defined by the user.  \cl is also capable of skimming and saving events using the auxiliary {\tt save} keyword in its internal format {\tt LVL0}, as described in Appendix~\ref{app:save}.  In case event saving is specified in the ADL file, the ROOT file also stores the saved events.

The output ROOT file includes a directory for each event categorization region, i.e. each {\tt region} block.  These directories contain all user-defined histograms specified in the ADL file.  The prototype version of \cl also had a basic {\tt cutflow} histogram listing the number of events surviving each step of the selection in the given region.  The cutflows, including the statistical errors on counts are also given as text output.  In the current version, the cutflow histograms are improved to include the selection criteria as bin labels.  Moreover, in case binning is used in a region, a {\tt bincounts} histogram is also added, where each histogram bin shows the event counts and errors in each selection bin, and the histogram bin labels show the bin definition.  The {\tt cutflow} and {\tt bincounts} histograms can be directly used in the subsequent statistical analysis of the results.  \blue{A screenshot of a simple example output can be seen in Figure~\ref{fig:histo-example} in Appendix~\ref{app:histo}.}

\subsection{Incorporation of existing counts}

In some cases, event counts and uncertainties from external sources are needed to be systematically accessible in order to be processed together with the counts and uncertainties obtained from running the analysis via \cl.  One example is phenomenological interpretation studies, where the analysis is only run through signal samples, while the experimental results, consisting of data counts and background estimates are usually taken from the experimental publication.  Having the data counts and background estimates directly available in a format compatible with the signal counts is necessary for subsequent statistical analysis.  Moreover, for this particular case, it is also highly desirable to have this information documented directly within the ADL file.  Another example is validation studies, when either multiple teams in an experimental group are synchronizing their cutflows, or a reimplemented analysis for a phenomenological interpretation study is validated against a cutflow provided by the original experimental publication.  Similarly, having the validation counts and uncertainties in the same format would make comparison very practical.  

Recently, a syntax was developed in ADL for systematically storing external counts and uncertainties within the ADL file. The physics process for which the information is given, and the format of the information is provided within the {\tt countsformat} block using the {\tt process} keyword, while the values are given in the relevant {\tt region} blocks right after the definition of the relevant selection criteria using the {\tt counts} keywords.  The syntax is detailed in Appendix~\ref{app:counts}.  When an ADL file including external counts and errors is run with \cl, the counts and errors are converted into {\tt cutflow} and {\tt bincounts} histograms with a similar format to those hosting the \cl output.  The histogram and are placed under the relevant region directories, and physics process is included in the histogram names.

\section{Performance and Multi-threaded runs}
\label{sec:multithread}

\begin{figure}[htbp]
\begin{center}
  \includegraphics[width=0.6\linewidth]{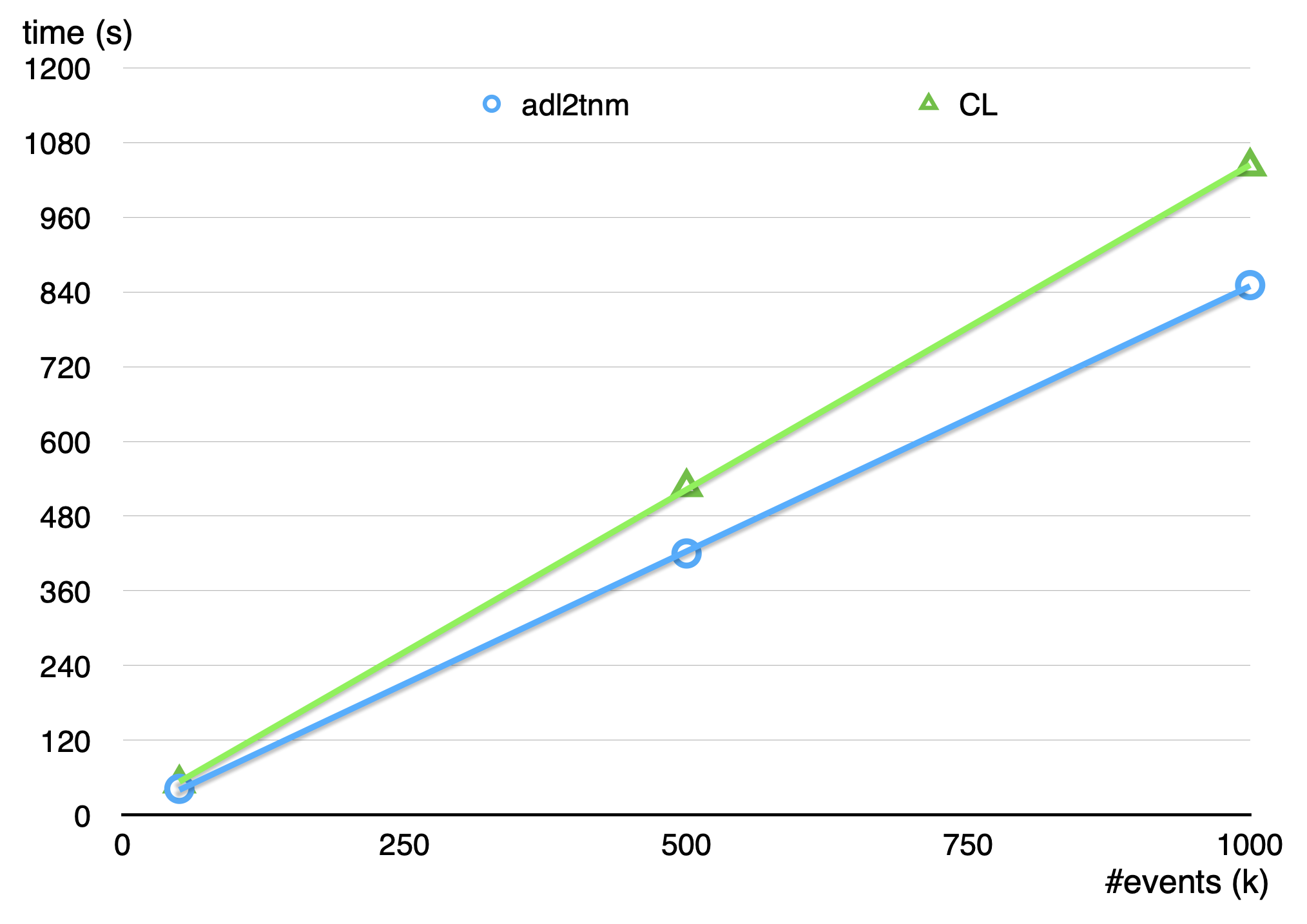}
  \caption{Speed comparison of \cl versus the C++ ADL transpiler {\tt adl2tnm} on a CMS supersymmetry analysis~\cite{Sirunyan:2018ell, adllhcanlrazor} using up to 1M supersymmetry signal events with the CMS NanoAOD format in a Mac OS setup.}
  \label{fig:speedcomparison}
\end{center}
\end{figure}

\blue{The \cl run-time interpreter is eventually aimed for use in the analysis of very large amounts of experimental data.  Therefore its speed and performance needs to be close to those of analyses tools based on GPLs.  It is expected that the process of run-time interpretation would decrease the performance due to additional tasks including lexical analysis, tokenization, etc.  Yet, at its current state, \cl's speed is only partially less than that of a C++ analyzer.  For a numerical test, a sufficiently complicated supersymmetry search analysis~\cite{Sirunyan:2018ell, adllhcanlrazor} involving multiple objects, 12 event categorization regions and several variable calculations based on external functions was run both with \cl and the C++-based ADL transpiler {\tt adl2tnm} using up to 1M supersymmetry signal events with the CMS NanoAOD format.  The speed comparison for running in a Mac OS setup is shown in Figure~\ref{fig:speedcomparison}.  Overall, \cl is about 20\% slower compared to the same analysis performed using a pure C++ code.   }

\cl has been also recently enhanced with the capability of multi-threaded execution of an analysis to optimally utilize the available resources and therefore get faster results. Adding \texttt{-j n} to the command to start the analysis execution enables using \texttt{n} number of cores, e.g. as
\begin{verbatim}
    ./CLA.sh [inputrootfile] [inputeventformat] -i [adlfilename] -j 2
\end{verbatim}
for 2 cores.  The requirement for \texttt{n} is to be an integer between \texttt{0} and the total number of cores on the processor, where the case of \texttt{-j 0} is used to select one less than the total number of cores to maximize performance for demanding analyses while leaving the operating system necessary part of the resources.  

\begin{figure}[htbp]
\begin{center}
  \includegraphics[width=0.7\linewidth]{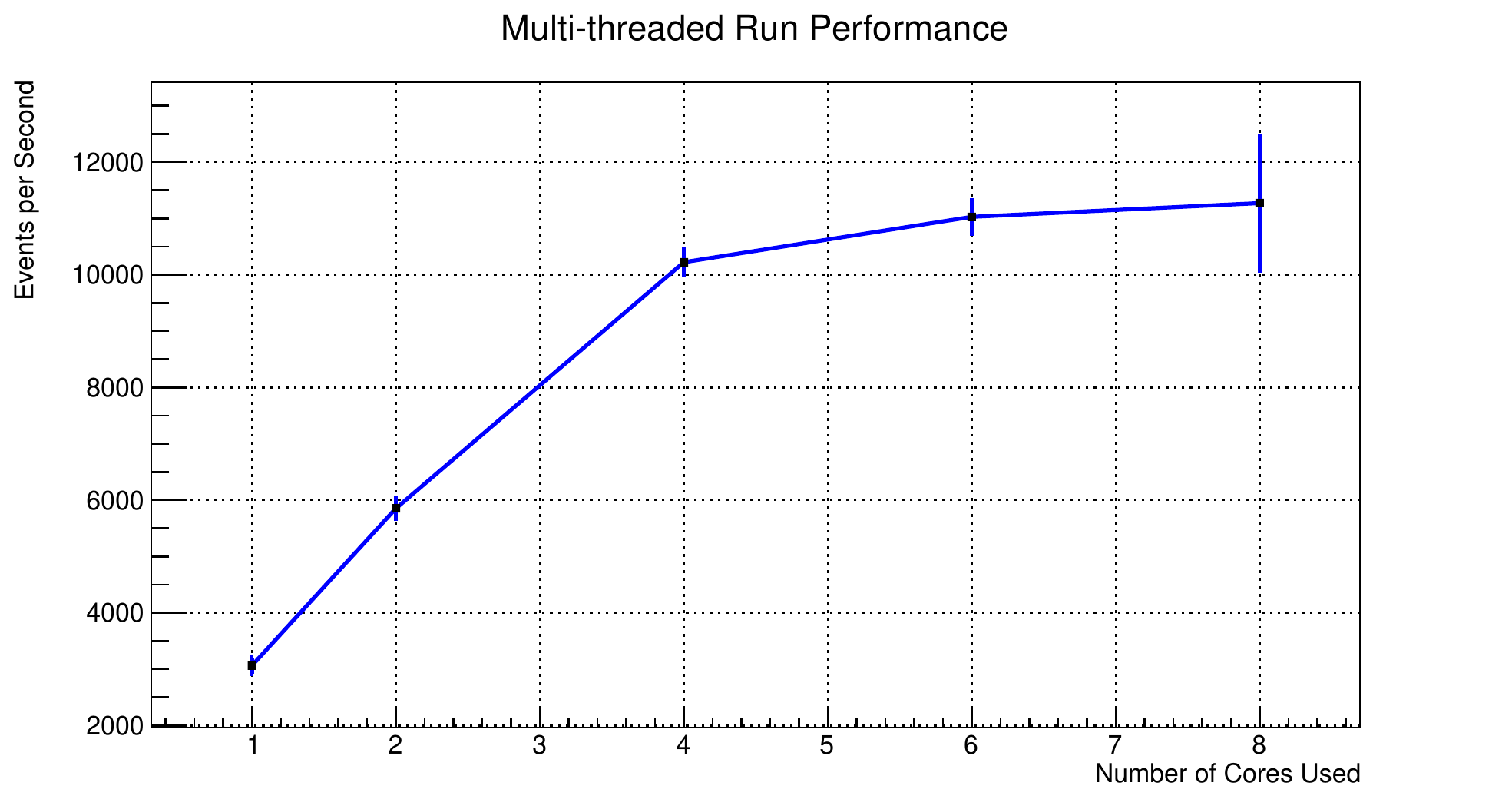}
  \caption{Events processed per second when analysis is divided into 1, 2, 4, 6 and 8 threads for varying number of events. Error bars are multiplied by 10 to make them visible.}
  \label{fig:parallel}
\end{center}
\end{figure}

Figure~\ref{fig:parallel} shows the run time dependence on multi-threading. The mean and standard deviation of these results are further given in Table~\ref{tab:fig_parallel}. The computer used during the test has Intel(R) Core(TM) i5-8300H with 4 cores, 8 threads and runs Ubuntu 18.04.4 LTS. 
\blue{ The number of events analyzed was limited to 3 million due to memory restrictions in the current ROOT implementation.
Although this is not the only possible way to collect results, it was convenient enough for a first implementation. It is surely possible to improve this implementation when the need arises by saving data on disk to free memory while continuing to run. 
}

\begin{table}[h]
\caption{Data points given in Figure \ref{fig:parallel}. \label{tab:fig_parallel}}

\centering{}%
\begin{tabular}{|c|c|c|}
\hline 
Threads & Mean no. of Events/sec  & Std.Dev. \tabularnewline
\hline 
\hline 
1 &  3063.4 & 14.5 \\
\hline 
2 & 5853.5 & 18.5 \\
\hline 
4 & 10223.3  & 22.3 \\
\hline 
6 & 11028.0  & 29.6 \\
\hline 
8 & 11272.0 & 119.6\\
\hline 

\end{tabular}
\end{table}

As can be seen from the results, total event processing rate increases linearly as the number of cores increase up to 4. Due to the processor having only 4 physical cores with 2 logical cores each, the runs that use more than 4 threads showed minimal improvement. Simultaneous processing efficiency, resource demand of background processes and recombination of results that are obtained in parallel also contribute to the decline in the multi-threaded run performance. 

In a different performance test, run times for 1,2,4 and 8 threaded analyses for varying numbers of events are given in Table~\ref{tab:plltime}. To simplify, a normalized version of Table~\ref{tab:plltime} is also provided in Table~\ref{tab:plltimenorm}, where the run time of an analysis that used a single core is taken to be the norm. Looking at these tables, it can be seen that, as the analyses get more complex, higher levels of multi-threading performance gets increasingly better.

\blue{A simple analysis task uses time mainly on reading data from disk and performing memory transfers. One should note that having a multicore system does not make an extra contribution in this scenario as there is only one disk. If the analysis becomes more complicated, the impact of read and copy operations gets reduced and CPU-intensive calculations start taking more time. Therefore in a CPU-intensive complex analysis, the benefit of having multiple cores becomes more pronounced.}

\begin{table}[h]
\caption{Variation of run times with changing number of threads.} \label{tab:plltime}
\centering{}%
\newcolumntype{D}{>{\centering\arraybackslash} p{3cm}}
\newcolumntype{C}{>{\centering\arraybackslash} p{2cm}}

\begin{tabular}{ |D|C|C|C|C|C|  }
\cline{3-6}
\multicolumn{1}{c}{}&\multicolumn{1}{c|}{} & \multicolumn{4}{ c| }{Process Time For Core Used [s]} \\ \cline{3-6}
\multicolumn{1}{c}{}&\multicolumn{1}{c|}{} & 1 & 2 & 4 & 8 \\ \cline{1-6}
\multicolumn{1}{ |c  }{\multirow{5}{*}{Processed Events} } &
\multicolumn{1}{ |c| }{${10^4}$} &   3.081       &   3.041   &   3.124       &   4.600     \\ \cline{2-6}
\multicolumn{1}{ |c  }{}                        &
\multicolumn{1}{ |c| }{${10^5}$}  &   21.085      &   12.062  &   8.316       &   9.630         \\ \cline{2-6} 
\multicolumn{1}{ |c  }{}                        &
\multicolumn{1}{ |c| }{${10^6}$}&   306.064     & 155.195   &   91.201      &   97.968     \\ \cline{2-6}
\multicolumn{1}{ |c  }{}                        &
\multicolumn{1}{ |c| }{${2.5\times10^6}$} &   776.133     & 402.723   &   227.817     &   209.623      \\ \cline{2-6}
\multicolumn{1}{ |c  }{}                        &
\multicolumn{1}{ |c| }{${4.5\times10^6}$} &   1409.416    & 722.901   &   416.964     &   374.946    \\ \cline{1-6}
\end{tabular}
\end{table}

\begin{table}[h]
\caption{Runtimes as percentages of single core runtime.} \label{tab:plltimenorm}
\centering{}%
\newcolumntype{D}{>{\centering\arraybackslash} p{3cm}}
\newcolumntype{C}{>{\centering\arraybackslash} p{2cm}}

\begin{tabular}{ |C|C|C|C|C|C|  }
\cline{3-6}
\multicolumn{1}{c}{}&\multicolumn{1}{c|}{} & \multicolumn{4}{ c| }{Normalized Process Time} \\ \cline{3-6}
\multicolumn{1}{c}{}&\multicolumn{1}{c|}{} & 1 & 2 & 4 & 8 \\ \cline{1-6}
\multicolumn{1}{ |c  }{\multirow{5}{*}{Processed Events} } &
\multicolumn{1}{ |c| }{${10^4}$} &   100  &  98.7  & 101 & 149    \\ \cline{2-6}
\multicolumn{1}{ |c  }{}                        &
\multicolumn{1}{ |c| }{${10^5}$}  &  100  &  57.2  & 38.6  & 45.7    \\ \cline{2-6} 
\multicolumn{1}{ |c  }{}                        &
\multicolumn{1}{ |c| }{${10^6}$}&   100  & 50.7 &  29.8 & 32.0 \\ \cline{2-6}
\multicolumn{1}{ |c  }{}                        &
\multicolumn{1}{ |c| }{${2.5\times10^6}$} &   100  &  51.9 & 29.4 & 27.0   \\ \cline{2-6}
\multicolumn{1}{ |c  }{}                        &
\multicolumn{1}{ |c| }{${4.5\times10^6}$} &   100  &  51.3  & 29.6  & 26.6  \\ \cline{1-6}
\end{tabular}
\end{table}

\section{Code maintenance and continuous integration}
\label{sec:maintenance}

The \cl source code is public and resides in the  popular software development platform GitHub~\cite{clgithub}: 

\vspace{0.2cm}
\href{https://github.com/unelg/CutLang}{https://github.com/unelg/CutLang}
\vspace{0.2cm}

\cl uses GitHub functionalities for parallel code development across multiple developers. This development platform, apart from a wiki page for documentation and possibility for error reporting, also offers a continuous integration setup which includes a series of tasks that could be initiated at a specific time or by a trigger such as a commit to the main branch.  The continuous integration setup was recently incorporated to automatically validate the code.  The setup compiles the \cl source code from scratch, and runs the resulting executable over a set of example ADL files from the package on a multitude of input data files and formats. 
\blue{By comparing the output from the examples to a reference output from earlier runs that were successfully executed and validated, any coding errors could be automatically detected and reported by email.}
The total compilation and execution time is greatly reduced by using a pre-compiled version of ROOT and by pre-installing the necessary event files onto a Docker~\cite{docker} image integrated to a recent Linux (Ubuntu) virtual computer made available by the development platform.

\section{Analysis examples
\label{sec:AnaEx} }

ADL and \cl are continuously being used for implementing a diverse set of LHC analyses and running these on events. The analyses implemented are being collected in the following GitHub repository~\cite{adllhcanl}:

\vspace{0.2cm}
\href{https://github.com/ADL4HEP/ADLLHCanalyses}{https://github.com/ADL4HEP/ADLLHCanalyses}
\vspace{0.2cm}

The main focus so far has been to implement analyses designed for new physics searches, in particular supersymmetry searches. These supersymmetry analyses are intended to be directly used to create model efficiency maps to be used by the reinterpretation framework SModelS~\cite{Kraml:2013mwa,Ambrogi:2017neo,Ambrogi:2018ujg}.  The results obtained by running some of the implemented analyses have also been validated within dedicated exercises performed during the Les Houches PhysTeV workshops, in comparison to other analysis frameworks~\cite{Brooijmans:2020yij}. The available analysis spectrum is currently being extended to cover Higgs and other SM analyses.  Furthermore, studies are ongoing to improve the functionalities of ADL and \cl for use in searches or interpretation studies with long-lived particles, which involve highly non-conventional objects and signatures.  More recently, analyses examples for CMS Open Data~\cite{opendata} and a sensitivity study case for High Luminosity LHC and the Future Circular Collider were also added~\cite{Paul:2020mul}.  In addition, ADL and \cl were used as main tools in an analysis school which took place in Istanbul in February 2020 for undergraduate students, and several analyses were implemented by the participating students~\cite{Adiguzel:2020brl}.  ADL and \cl were also used to prepare hands-on exercises for data analysis at the 26th Vietnam School of Physics (VSOP) in December 2020~\cite{vsop}. The VSOP exercises involving running \cl and further analysis of resulting histograms with ROOT were  also adapted for direct use via Jupyter notebooks, and are documented in detail in~\cite{vsophandson}.  The experience in both schools justified ADL and \cl as highly intuitive tools for introducing high energy physics data analysis to undergraduate and masters students with nearly no experience in analysis.

Implementing analyses with a variety of physics content led to incorporating a wider range of object and selection operations and helped to make the ADL syntax more generic and inclusive. Syntax for generalizing object combinations, numerical efficiency applications, hit-and-miss method, bins and counts and many others were introduced as a result of these studies. Consequently, the scope and functionality of \cl interpreter and framework was also enhanced.  Many internal and external functions were added to \cl to address direct requirements of the various implemented analyses.  Running different analyses on events also allowed to thoroughly test the capacity of \cl in performing complete, realistic analysis tasks.

\section{Conclusions}
\label{sec:conclusions}

We presented the recent developments in \cl, leading towards a more complete analysis description language and a more robust runtime interpreter.  The original syntax of the earlier \cl prototype version and its event processing methods have been modified after a multitude of discussions with other scientists in the field interested in decoupling the physics analysis algorithms from the computational details and after implementing many HEP analyses. Modifications include significant enhancement of object definition and event classification expressions, addition of more functions for calculating event variables, incorporation of tables for applying efficiencies, adaptation of a system for including external counts, and more.  Although these modifications broke the strict backward compatibility of the earlier version of the language, we believe they should be considered as improvements as they certainly will lead to a cleaner, more robust and a widely accepted analysis description language. The improved syntax processing relies on formal lexical and grammar definition tools widely available in all Unix-like operating systems.

One direct result of the syntax modifications originating from community-wide discussions is that, in the presented version there are more than a single way of expressing the same idea in \cl. We believe this is a desirable property: after all, in human languages (that we try to imitate) as well, the same idea can be expressed in multiple ways. To give an example to reject events with a property smaller than a certain threshold amounts to accepting events greater than the same threshold. Such a property should not be considered as a source of potential confusion and error, but as a fertility of the language.

\cl still follows the approach of runtime interpretation.  We strongly believe that direct interpretation of the human readable commands and algorithms, although slower in execution as compared to a compiled binary, leads to faster and less error-prone algorithm development. The possible event processing speed issues can be cured by parallel processing of independent events and regions. The interpreted and human readable nature of \cl and ADL have a potential area of growth and development: with the advance of machine learning hardware and software tools, the dream of being able to perform an LHC-type analysis just by talking to the computer in one's native tongue might not be too far-fetched.

\blue{The advances described in this paper brought ADL and \cl to a state where they can handle many standard analysis expressions and operations and have developed the earlier prototype into a practically usable infrastructure. \cl at its current stage can directly perform phenomenological studies and some simple experimental studies.  However there are still some limitations to address in the language and the interpreter.  In the near future, ADL syntax will be further expanded by inclusion of a generic way to describe arbitrary combinations of objects to form new ones, the capability of adding new object attributes and defining object associations, lower level objects or non-standard objects such as long-lived particles.  One major addition would be the capability to express and handle variations due to systematic uncertainties.  Moreover, the interpreter would benefit from further automatizing the incorporation of new input data types or external functions, which currently require manual intervention from the users.  Enabling an automated syntax verification and providing explicit guidance for possible syntax errors would further facilitate the analysis process.   Plans are underway to improve the design of the \cl infrastructure in the near future based on current best practices in compiler construction to accommodate all these features and arrive at a more robust, yet flexible and user-friendly analysis ecosystem.  With the growing data, our field will undoubtedly continue conceiving new analysis concepts and methods which may not be immediately applicable in ADL and \cl.  The current developer team is dedicated to following and implementing these features.  Yet, we foresee that the planned improvements in the fundamental design of ADL and \cl will lead the progress towards the ultimate goal of analysis automation.}

Finally, as any language, \cl/ADL grows with the people that use it to solve new problems. With every analysis requiring a new functionality, the list of already-solved problems grows. We hope that, such an internal library together with the script assisted addition of external user functions will allow the analysts of the future to spend less time on previously solved problems and to focus their energy in innovating  solutions to the analysis problems of the post LHC era colliders.

\section*{Acknowledgements}

We thank Harrison B. Prosper for useful discussions on the language content and help with validation of analysis results.  We also thank the SModelS team for a collaboration that is helping to gradually improve CutLang.  SS is supported by the National Research Foundation of Korea (NRF), funded by the Ministry of Science \& ICT under contracts 2021R1I1A3048138 and NRF-2008-00460.

\bibliographystyle{JHEP_mod}
\bibliography{refs}{}
\newpage

\appendix
\section{User Manual}
\label{appendix:A}
\renewcommand{\thesubsection}{A.\arabic{subsection}}

All information about ADL and \cl including publications, talks and twikis with syntax rules can be accessed through the following portal

\vspace{0.2cm}
\href{https://cern.ch/adl}{https://cern.ch/adl}
\vspace{0.2cm}

\noindent The code for \cl is hosted in the GitHub repository

\vspace{0.2cm}
\href{https://github.com/unelg/CutLang}{https://github.com/unelg/CutLang}
\vspace{0.2cm}

\noindent which provides up-to-date instructions on how to install, compile and run \cl.

\subsection{Blocks and keywords}

An ADL file consists of blocks based on a keyword value/expression structure.  The blocks allow a clear separation of analysis components.  A typical block looks as follows:

\begin{verbatim}
blockkeyword blockname
    # general comment
    keyword1 expression1
    keyword2 expression1
    keyword3 expression1 # comment about value3
\end{verbatim}

Table~\ref{tab:blocks} lists the available blocks, their purposes and associated keywords, and Table~\ref{tab:keywords} lists the keywords.  The details on their applications are given in the following sections.

\begin{table}[h]
\caption{Blocks in ADL and \cl 
\label{tab:keywords}}
\centering{}%
\begin{tabular}{|p{0.20\textwidth}|p{0.6\textwidth}|p{0.15\textwidth}|}
\hline 
Block & Purpose & Related keywords \\
\hline 
\hline 
object / obj & Object definition block. Produces an object type from an input object type by applying selections. & take, select, reject \\
\hline
region / algo & Event categorization. & select, reject, weight, bin, sort, counts, histo, save \\
\hline
info & Contains analysis information such as the experiment, center-of-mass energy, luminosity, publication details, etc. & \\
\hline
table & Generic block for tabular information, such as efficiency values versus variable ranges & tabletype, nvars, errors \\
\hline 
countformat & Expresses the processes for which external counts are included and the format of counts & process \\
\hline
\end{tabular}
\end{table}

\begin{table}[h]
\caption{Keywords in ADL and \cl 
\label{tab:blocks}}
\centering{}%
\begin{tabular}{|p{0.20\textwidth}|p{0.6\textwidth}|p{0.15\textwidth}|}
\hline 
Keyword & Purpose & Related block \\
\hline 
\hline 
define & Define variables, constants & -- \\
\hline
select & Select objects or events based on criteria that follow the keyword. & object, region \\
\hline
reject & Reject objects or events based on criteria that follow the keyword. & object, region \\
\hline
take / using / : & Define the mother object type & object \\
\hline 
sort & Sort an object in an ascending or descending order wrt a property. & region \\
\hline
weight & Weight events & region \\
\hline
histo & Fill histograms & region \\
\hline
process & Specify process and the format for which external counts are given & countformat \\
\hline
counts & Give external counts & region \\
\hline
tabletype & Specifies type of the table & table \\
\hline
nvars & Number of variables in a table & table \\
\hline
errors & Type of errors indicated in a table & table \\
\hline
title, experiment, id, publication, sqrtS, lumi, arXiv, hepdata, doi & Provide information about the analysis (see Table~\ref{tab:info-section}) & info \\
\hline
\end{tabular}
\end{table}

\subsection{Predefined physics objects}
\label{app:predefobj}

Basic physics objects and their properties currently available in \cl are defined in Table~\ref{tab:Basic-physics-objects}. The predefined particles are initially sorted per decreasing transverse momentum and their indices start at zero. With the current implementation, all the predefined particle names, and commonly used function names have become case-insensitive. For the particle, both Python-type and \LaTeX-type notations are accepted; the former with square brackets, and the latter with an underline character. An example for electrons is given below:
\begin{verbatim}
     Ele_0  = ELE_0 = Ele[0] = ele[0] = electron_0 = electron[0]  .
\end{verbatim}

Sometimes it is necessary to refer to the whole object set or just to some of its members. The \cl notation for these cases is to write the name of the set without any indices for the former (i.e. \texttt{ELE} ) and to use the semi-colon notation for the latter (i.e. \texttt{ELE[0:2] = ELE\_0:2} ) .

In \cl, there are two object-types that merit special attention: the lepton and the neutrino types. The \texttt{LEP} keyword refers to a generic lepton and at runtime it is reduced to an electron or to a muon depending on the choice as explained in Table \ref{tab:initializations}. This helps the physicist avoiding two algorithm sections, one for electron and other muon based analyses.  The second object-type is related to the taming of the neutrino escaping from the detector. At LHC energies and beyond, for which \cl is intended, the W bosons are generally produced with a sufficient boost such that in the leptonic decays, the pseudorapidity of the charged lepton is not very different from the chargeless one. Therefore this particular physics object benefits from this approximation to define a massless and chargeless particle with transverse momentum and azimuthal angle ($\phi$) values extracted from the missing transverse energy (MET) measurements. The pseudorapidity, however, is taken equal to that of the charged lepton with the same particle index.

\begin{table}[h]
\caption{Basic physics object nomenclature in \cl \label{tab:Basic-physics-objects}}

\centering{}%
\begin{tabular}{|r|l|c|c|c|c|c|}
\hline 
Name & Keyword & \multicolumn{2}{c|}{First object} & \multicolumn{2}{c|}{Second object} & $j+1^{th}$ object\tabularnewline
\hline 
\hline 
Electron & \texttt{ELE} & \texttt{ELE{[}0{]}} & \texttt{ELE\_0} & \texttt{ELE{[}1{]}} & \texttt{ELE\_1} & \texttt{ELE\_j}\tabularnewline
\hline 
Muon & \texttt{MUO} & \texttt{MUO{[}0{]}} & \texttt{MUO\_0} & \texttt{MUO{[}1{]}} & \texttt{MUO\_1} & \texttt{MUO\_j}\tabularnewline
\hline 
Tau & \texttt{TAU} & \texttt{TAU{[}0{]}} & \texttt{TAU\_0} & \texttt{TAU{[}1{]}} & \texttt{TAU\_1} & \texttt{TAU\_j}\tabularnewline
\hline 
Lepton & \texttt{LEP} & \texttt{LEP{[}0{]}} & \texttt{LEP\_0} & \texttt{LEP{[}1{]}} & \texttt{LEP\_1} & \texttt{LEP\_j}\tabularnewline
\hline 
Photon & \texttt{PHO} & \texttt{PHO{[}0{]}} & \texttt{PHO\_0} & \texttt{PHO{[}1{]}} & \texttt{PHO\_1} & \texttt{PHO\_j}\tabularnewline
\hline 
Jet & \texttt{JET} & \texttt{JET{[}0{]}} & \texttt{JET\_0} & \texttt{JET{[}1{]}} & \texttt{JET\_1} & \texttt{JET\_j}\tabularnewline
\hline 
Fat Jet & \texttt{FJET} & \texttt{FJET{[}0{]}} & \texttt{FJET\_0} & \texttt{FJET{[}1{]}} & \texttt{FJET\_1} & \texttt{FJET\_j}\tabularnewline
\hline 
b-tagged Jet & \texttt{BJET} & \texttt{BJET{[}0{]}} & \texttt{BJET\_0} & \texttt{BJET{[}1{]}} & \texttt{BJET\_1} & \texttt{BJET\_j}\tabularnewline
\hline 
light Jet & \texttt{QGJET} & \texttt{QGJET{[}0{]}} & \texttt{QGJET\_0} & \texttt{QGJET{[}1{]}} & \texttt{QGJET\_1} & \texttt{QGJET\_j}\tabularnewline
\hline 
Neutrino & \texttt{NUMET} & \texttt{NUMET{[}0{]}} & \texttt{NUMET\_0} & \texttt{NUMET{[}1{]}} & \texttt{NUMET\_1} & \texttt{NUMET\_j}\tabularnewline
\hline 
MET & \texttt{METLV} & \texttt{METLV{[}0{]}} & \texttt{METLV\_0} & --- & --- & ---\tabularnewline
\hline 
generator particle & \texttt{GEN} & \texttt{GEN{[}0{]}} & \texttt{GEN\_0} & \texttt{GEN{[}1{]}} & \texttt{GEN\_1} & \texttt{GEN\_j}\tabularnewline
\hline 
\end{tabular}
\end{table}

\subsection{Predefined functions}
\label{app:predeffunc}

Functions in \cl can be used for accessing object attributes, or for computing new variables from object or event quantities.  Functions for accessing object attributes can be directly related to Lorentz vectors such as mass, momentum, rapidity etc, or be related to other variables found in some commonly used ntuples. In both cases, both the function syntax with parentheses and the attribute syntax with curly braces can be used.  Functions used for computing new quantities can use object attributes or other already calculated quantities or constants.  The currently available object attribute functions in \cl are listed in Table~\ref{tab:objfunctions}. Note that some of the attributes listed here are only valid for certain input types, e.g. for CMS NanoAOD, but not for others, e.g. for Delphes.  The functions used for computing new quantities are listed in Table~\ref{tab:newquantityfunctions}. 

One should note that in \cl adding particles could be achieved by either writing these one after the other separated by space(s), or by using a $+$ sign. Both notations are equally valid. Additionally, one should use a comma as the separation character for the functions requiring multiple arguments.

The internal functions, such as angular distance or transverse momentum are also case-insensitive in \cl, though they are written in this manuscript with a certain syntax (first letter upper case) for clarity in reading. The functions requiring multiple arguments should use comma character for argument separation. External functions can also be downloaded and added to \cl library. The instructions for this operation is described in appendix \ref{appendix:B}.

\begin{table}[h]
\caption{Functions and syntax for object attributes in \cl.} \label{tab:objfunctions}
\centering{}%
\begin{tabular}{|r|r||l|}
\hline 
\textbf{Meaning} & Syntax 1 & Syntax 2\tabularnewline
\hline 
\multicolumn{3}{|c|}{\it Lorentz vector-related attributes} \\
\hline 
Mass of & \texttt{m( )} & \texttt{\{ \}m}\tabularnewline
\hline 
Charge of & \texttt{q( )} & \texttt{\{ \}q}\tabularnewline
\hline 
Phi of & \texttt{Phi( )} & \texttt{\{ \}Phi}\tabularnewline
\hline 
Eta of & \texttt{Eta( )} & \texttt{\{ \}Eta}\tabularnewline
\hline 
Absolute value of Eta of & \texttt{AbsEta( )} & \texttt{\{ \}AbsEta}\tabularnewline
\hline 
Rapidity of & \texttt{Rep( )} & \texttt{\{ \}Rep}\tabularnewline
\hline 
Pt of & \texttt{Pt( )} & \texttt{\{ \}Pt}\tabularnewline
\hline 
Pz of & \texttt{Pz( )} & \texttt{\{ \}Pz}\tabularnewline
\hline 
Energy of & \texttt{E( )} & \texttt{\{ \}E}\tabularnewline
\hline 
Momentum of & \texttt{P( )} & \texttt{\{ \}P}\tabularnewline
\hline 
\multicolumn{3}{|c|}{\it Other attributes} \\
\hline 
PDGID of a particle & \texttt{PDGID( )} & \texttt{\{ \}PDGID}\tabularnewline
\hline 
Charge of a particle & \texttt{btagDeepB( )} & \texttt{\{ \}btagDeepB}\tabularnewline
\hline 
is the jet b tagged? & \texttt{bTag( )} & \texttt{\{ \}bTag}\tabularnewline
\hline 
Soft Drop mass of a jet & \texttt{msoftdrop( )} & \texttt{\{ \}msoftdrop}\tabularnewline
\hline 
N-subjetiness variable 1 & \texttt{tau1( )} & \texttt{\{ \}tau1}\tabularnewline
\hline 
N-subjetiness variable 2 & \texttt{tau2( )} & \texttt{\{ \}tau2}\tabularnewline
\hline 
N-subjetiness variable 3 & \texttt{tau3( )} & \texttt{\{ \}tau3}\tabularnewline
\hline 
Leptonic diTau invariant mass & \texttt{fMTauTau( )} & \texttt{\{ \}fMTauTau}\tabularnewline
\hline 
transverse impact parameter & \texttt{dxy( )} & \texttt{\{ \}dxy}\tabularnewline
\hline 
longitudinal impact parameter & \texttt{dz( )} & \texttt{\{ \}dz}\tabularnewline
\hline 
lepton identification variable  & \texttt{softId( )} & \texttt{\{ \}softId}\tabularnewline
\hline 
relative isolation for leptons  & \texttt{miniPFRelIsoAll( )} & \texttt{\{ \}miniPFRelIsoAll}\tabularnewline
\hline
 MVA based tau ID  & \texttt{dMVAnewDM2017v2( )} & \texttt{\{ \}dMVAnewDM2017v2}\tabularnewline
\hline 
$\sigma_{i\eta i\eta}$ for photons & \texttt{sieie( )} & \texttt{\{ \}sieie} \\
\hline 
isolation variable & \texttt{reliso( )} & \texttt{\{ \}reliso} \\
\hline 
isolation variable & \texttt{relisoall( )} & \texttt{\{ \}relisoall} \\
\hline 
isolation variable & \texttt{pfreliso03all( )} & \texttt{\{ \}pfreliso03all} \\
\hline 
Tau decay mode id & \texttt{iddecaymode( )} & \texttt{\{ \}iddecaymode} \\
\hline 
Tight ID and isolation flag & \texttt{idisotight( )} & \texttt{\{ \}idisotight} \\
\hline 
Tight anti ele ID for taus & \texttt{idantieletight( )} & \texttt{\{ \}idantieletight} \\
\hline 
Tight anti mu ID for taus & \texttt{idantimutight( )} & \texttt{\{ \}idantimutight} \\
\hline 
Tight ID for muons & \texttt{tightid( )} & \texttt{\{ \}tightid} \\
\hline 
PU ID for jets & \texttt{puid( )} & \texttt{\{ \}puid} \\
\hline 
Index of matched genparticle to a lepton & \texttt{genpartidx( )} & \texttt{\{ \}genpartidx} \\
\hline 
Tau decay mode & \texttt{decaymode( )} & \texttt{\{ \}decaymode} \\
\hline 
Tau isolation & \texttt{tauiso( )} & \texttt{\{ \}tauiso} \\
\hline 
Muon soft ID & \texttt{softId( )} & \texttt{\{ \}softId} \\
\hline 
\end{tabular}
\end{table}

\begin{table}[h]
\caption{Functions and syntax for computing new quantities in \cl. 
\label{tab:newquantityfunctions}}
\centering{}%
\begin{tabular}{|r|r||l|}
\hline 
\textbf{Meaning} & Syntax 1 & Syntax 2\tabularnewline
\hline
Angular distance between & \texttt{dR( )} & \texttt{\{ \}dR}\tabularnewline
\hline 
Phi difference between & \texttt{dPhi( )} & \texttt{\{ \}dPhi}\tabularnewline
\hline 
Eta difference between & \texttt{dEta( )} & \texttt{\{ \}dEta}\tabularnewline
\hline 
Missing transverse energy in the event & \texttt{MET} & --\tabularnewline
\hline 
sum of jet transverse momenta & \texttt{HT( )} & --\tabularnewline
\hline
partitioning objects into 2 megajets & \texttt{fmegajets( )} & \texttt{\{ \}fmegajets}\tabularnewline
\hline 
Razor variable MR & \texttt{fMR( )} & \texttt{\{ \}fMR}\tabularnewline
\hline 
Razor variable MTR & \texttt{fMTR( )} & \texttt{\{ \}fMTR}\tabularnewline
\hline 
\hline 
partitioning objects into 2 hemispheres & \texttt{fhemisphere( )} & {\texttt{\{ \}fhemisphere}} \tabularnewline
\hline 
transverse mass MT2 & \texttt{fMT2( )} & \texttt{\{ \}fMT2} \tabularnewline
\hline 
\end{tabular}
\end{table}

\subsubsection{PDGID of particles}
\label{app:pdgid}

Each type of particle recognized in particle physics is assigned a unique code by the Particle Data Group (PDG) in order to facilitate interface between event generators, detector simulators, and analysis packages. These codes are known as PDGID (or PDG ID), and this method is called the MC particle numbering scheme~\cite{pdgid}. The numbering includes elementary particles such as, electrons, neutrinos, Z bosons etc, composite particles (mesons, baryons etc) and atomic nuclei. Hypothetical particles beyond the Standard Model also have PDGIDs. Particles have a positive PDGID whereas antiparticles a negative one. The list of PDGID of some particles is given in table \ref{table:pdgid}

\begin{table}[h!]
\caption{PDGID of some elementary particles\cite{PDGreview}}
\centering
\begin{tabular}{|ll|ll|ll|}
\hline
\multicolumn{2}{|l|}{Quarks} & \multicolumn{2}{l|}{Leptons} & \multicolumn{2}{l|}{Bosons} \\ \hline
d             & 1            & $e^-$               & 11         & $\gamma$        & 22        \\ \hline
u             & 2            & $\mu^-$           & 13         & Z               & 23        \\ \hline
s             & 3            & $\tau^-$          & 15         & $W^+$               & 24        \\ \hline
\end{tabular}
\label{table:pdgid}
\end{table}

\cl provides an internal function that obtains a particle's PDGID.  Particles of a certain type can be selected using this functionality, e.g. :
\begin{verbatim}
    select  PDGID( LEP[0]) == -11
\end{verbatim}
This command selects positrons. (Positron is the antiparticle of electron, therefore it has a negative PDGID)

\subsection{Mathematical operators and functions}

Mathematical functions available in \cl are listed in Table~\ref{tab:mathfunc}.  
Trigonometric and logarithmic functions are implemented with their usual meanings. The Heaviside step function or the unit step function {\tt hstep}, which was also added recently, is a discontinuous function, named after Oliver Heaviside, whose value is zero for negative arguments and one for positive arguments. 
The reducer functions for minimization and maximization, {\tt min} and {\tt max}, which were added recently, are discussed in Appendix~\ref{app:maxmin}. The reducer function {\tt size} $/$ {\tt count} returns the number of elements of a given set, such as the number of electrons.

\begin{table}[H]
\caption{mathematical and logical operators}
\label{tab:mathfunc}
\centering{}%
\begin{tabular}{|r|l||r|l|}
\hline 
Meaning & Operator & Meaning & Operator\tabularnewline
\hline 
\hline 
number of & Size( ) Count() NumOf() & absolute value & abs()\tabularnewline \hline 
tangent & tan() & hyperbolic tangent & tanh()\tabularnewline \hline 
sine & sin() & hyperbolic sine & cosh()\tabularnewline \hline 
cosine & cos() & hyperbolic cosine & sinh()\tabularnewline \hline 
natural exponential & exp() & natural logarithm & log()\tabularnewline \hline 
square root & sqrt() & Heaviside step function & hstep()\tabularnewline \hline 
as close as possible & \textasciitilde = & usual meaning & + - / {*}\tabularnewline \hline 
as far away as possible & \textasciitilde ! & to the power & \textasciicircum\tabularnewline \hline 
\end{tabular}
\end{table}

\subsection{Comparison, range and logical operators}
\label{app:comprangelog}

\cl understands the basic mathematical comparison expressions and logical operations. \texttt{ C/C++} operator notations and their Fortran counterparts are recognized and correctly interpreted. Additionally square brackets are used to define inclusive or exclusive ranges. The available comparison, range and logical operators can be found in Table \ref{tab:comprangelog}.

\begin{table}[h]
\caption{Comparison, range and logical operators in \cl\label{tab:comprangelog}}

\centering{}%
\begin{tabular}{|c|l|}
\hline 
Keywords & Explanation\tabularnewline
\hline  \hline 
\texttt{> >= == <= < } & usual meaning\tabularnewline \hline 
\texttt{GT GE EQ LE LT } & usual meaning\tabularnewline \hline 
\texttt{!= NE} & not equal\tabularnewline \hline 
\texttt{[ ]} & in the interval\tabularnewline \hline 
\texttt{{]} {[}} &  not in the interval \tabularnewline \hline 
\texttt{NOT} & logical not\tabularnewline \hline 
\texttt{AND and \&\&} & logical and\tabularnewline \hline 
\texttt{OR or ||} & logical or\tabularnewline \hline 
\end{tabular}
\end{table}

\subsubsection{Logical operations}
The use of Boolean operators (AND, OR, NOT) can make it easy to write the event selection criteria. In \cl, logical AND and logical OR operator had already been used to combine multiple event selection criteria. The newly implemented logical NOT simplifies the way to write the criteria of event selections in the analysis code to a great extent. The simplest example code to understand the syntax:
\begin{verbatim}
    select NOT Size(ELE) > 4
\end{verbatim}
This command selects events which do NOT have number of electrons greater than 4. However, the advantage of the NOT operator becomes more apparent when trying to negate more complex selections.  The event selection criteria can be combined using the logical AND, OR, NOT. For example :
\begin{verbatim}
    select (NOT condition1 ) AND ( condition2 OR condition3 )
\end{verbatim}
Now let us look at another code : 
\begin{verbatim}
    select Size(ELE) == 2 
    select NOT ( {ELE[0] ELE[1]}q == 0 AND {ELE[0] ELE[1]}m [] 80 100)
\end{verbatim}
The criteria \codeword{( {ELE[0] ELE[1]}q == 0 AND {ELE[0] ELE[1]}m [] 80 100)} can be used for defining Z bosons. As we have set NOT, we veto events with Z boson while looking for other dilepton signatures. Without using the \codeword{NOT} command, this selection would not be so straightforward, and would require a more complicated expression.

\subsubsection{Ternary operator}

Application of conditional selection criteria is available, including nested statements, using a syntax similar to that of C++ :
\begin{verbatim}
    condition ? true-case : false-case
\end{verbatim}

The following example illustrates a use case: if the number of \texttt{muonsVeto} particles equals to 1, then the \texttt{MTm} quantity should be less than 100 otherwise the \texttt{MTe} quantity should be less than 100:

\begin{verbatim}
Size(muonsVeto) == 1 ? MTm < 100 : MTe < 100 
\end{verbatim}

\subsection{$\chi^{2}$ minimization 
\label{app:chi2min}}

In an analysis with a multitude of objects of the same type, the analyst could search for the best combination defined by some criterion. A typical example, used in fully hadronic $t\bar{\ensuremath{t}}$ reconstruction would be to find the jet combination that would yield the best $W$ boson mass, or to find the two charged leptons that would result in the best $Z$ boson mass. The need for such a search can be expressed in \cl using two special comparison operators: \texttt{\textasciitilde =} and \texttt{\textasciitilde !}. The former is used in the sense of ``as close as possible to" whereas the latter for calculation ``as far as possible from". These two operators can be used to express $\chi^{2}$ minimization kinds of operations. The indices of the particles in such a search are to be given as negative. For example, the statement ``find two leptons with a combined invariant mass as close to 90.1 GeV" can be expressed in \cl notation as \texttt{\{ LEP\_-1 LEP\_-1 \}m \textasciitilde = 90.1} . In this case, \cl finds the best pair of particles satisfying the condition, and stores it per event for possible later use. However the analyzer should not use negative indices directly inside the {\tt region} block. It is a much better practice that improves readability to define a new object such as\texttt{ define ZLepRec = LEP{[}-1{]} LEP{[}-1{]}}. This definition can be used when defining histograms or other selection criteria, such as when selecting the charge of the found lepton pair, etc. If another particle of the same type (e.g. another lepton) is to be found, it is necessary to use a different but still negative index value. 


\subsection{Definitions}
ADL and \cl allow to assign alias names to constants (e.g. Z boson mass) or variables (e.g. angular variables between objects, mass of the Z boson reconstructed from two leptons, etc.).  The syntax and examples are given in Table~\ref{tab:Simple-definitions}.  Note that the keyword \texttt{define} can also be shortened as \texttt{def}.

\begin{table}[H]
\caption{Simple definitions\label{tab:Simple-definitions}}
\centering{}%
\begin{tabular}{|c|c|c|c|l|}
\hline 
Keywords & argument1 & symbol\tablefootnote{both : and = can be used interchangeably} & argument1 & Example\tabularnewline
\hline 
\hline 
\texttt{define} & name  & :/= & value & define mZprime = 500 \tabularnewline
\hline 
\texttt{define} & name  & :/= & function & define mTop1 : m(Top1) \tabularnewline
\hline 
\texttt{define} & name & :/= & particle(s) & define Zreco : ELE{[}0{]} ELE{[}1{]} \tabularnewline
\hline 
\end{tabular}
\end{table}

\subsection{Tables}
\label{app:tables}

The present version of \cl incorporates tables to implement various HEP related quantities, such as efficiencies, acceptances or trigger turn-on curves. Currently only one and two-dimensional tables can be used. These tables should have a name and a table type, specified by the {\tt tabletype} keyword, where the latter defines what information is hosted by the table.  Currently, only efficiency tables are recognized, therefore the table type information only serves as documentation and is not used by the interpreter.  However, as other uses for tables are developed, table type would become more relevant in the future.  Tables must also specify the number of variables (1 or 2) using the {\tt nvars} keyword as well as the availability of errors on the central value (true or false) using the {\tt errors} keyword. These should be followed by the table data, using the value [lower-error upper-error] lower-limit1 upper-limit1 [lower-limit2 upper-limit2] notation. Once defined in the definitions section, the table can be referred to and used in object and event selection.  An example table is shown below:
\begin{verbatim}
table tightmuoneff
      tabletype efficiency
      nvars 2
      errors true
#     value   err-  err+      min    max        min    max   
        0.1   0.01  0.02       0.0   10.0      -5.5    0.0
        0.1   0.01  0.02       0.0   10.0       0.0    5.5
        0.2   0.01  0.03      10.0   20.0      -5.5    0.0
        0.2   0.01  0.03      10.0   20.0       0.0    5.5
        0.4   0.01  0.04      20.0   50.0      -5.5    0.0
        0.4   0.01  0.04      20.0   50.0       0.0    5.5
        0.7   0.01  0.05      50.0   70.0      -5.5    0.0
        0.7   0.01  0.05      50.0   70.0       0.0    5.5
        0.95  0.01  0.06      70.0   1000.0    -5.5    0.0
        0.95  0.01  0.06      70.0   1000.0     0.0    5.5
\end{verbatim}

\subsection{Manipulating objects}

\subsubsection{Defining new objects}
New objects can be declared using a simple syntax:
\begin{verbatim}
 object new_object_name : base_object_name
\end{verbatim}
where the \texttt{object} keyword can also be shortened as \texttt{obj}, and instead of the symbol \texttt{:}, the keywords \texttt{using}  and \texttt{take} can be used.  The base object name can be a base object class, or a previously defined new object type such in the case of defining b-tagged jets from already defined high transverse momentum jets. These are usually called derived objects. An example, defining a derived new electron type based on predefined electrons would be written as:
\begin{verbatim}
 obj goodEle : ELE 
\end{verbatim}



One way of defining a derived object type is to list a set of selection critieria that distinguishes it from the base object, such as:

\begin{verbatim}
object AK4jets 
    take JET 
    select {JET_}Pt > 30 
    select {JET_}AbsEta < 2.4 
\end{verbatim}

It is also possible to create a new object by forming a group out of multiple base or derived objects, for example, to create a lepton object from electrons and muons. This is achieved using the \texttt{Union} function, as shown below.  This particular case of new object creation does not use any selection.

\begin{verbatim}
object leps : Union( MUO , ELE, TAU) # add all leptons into a set 

object gleps : Union( goodEle , goodMuo ) # add all derived leptons into another set
\end{verbatim}

\subsubsection{Sorting objects}
\label{app:sort}

By default, objects are sorted according to their transverse momentum, $p_t$, in descending order. 
For example, \codeword{ELE[0]} denotes the electron having the highest transverse momentum.  In some cases, objects may need to be sorted according to some other property, such as energy, pseudorapidity etc. In the current version, this can be done as:
\begin{verbatim}
    sort {ELE_ }E ascend
\end{verbatim}
This command sorts electrons according to their energy in ascending order, i.e. \codeword{ELE[0]} will have the least energy.  Sorting can also be done in the descending order by using {\tt descend}.

\subsubsection{Object combinatorics}
\label{app:comb}

Let us assume that we have an event with 5 jets, and we would like to reconstruct all hadronic Z bosons in the event.  What are the combinations? Numbering the jets from 1 to 5, some possibilities are given in Table ~\ref{tab:Combination-example}, in the left panel. It is obvious that not all possibilities are listed, and finally only one possibility can be true: after all a jet can not be used to reconstruct two different Z bosons. On top of this, other requirements might be applied to further restrict the possible Z candidates. For example, there might be a pseudorapidity range limit on each candidate, the transverse momentum of the jets forming the Z boson could be limited, the angular separation between the hadronic Z candidate and the first constituent jet might be limited, and finally, the invariant mass of the Z candidate might be requested to be in a certain range. After all these restrictions, the same initial set might be reduced to the combinations listed in the right panel of  Table~\ref{tab:Combination-example}, where the candidates that did not pass the requirements are shown as stroked out.

\begin{table}[H]
\caption{Combining two jets to reconstruct a hadronic Z boson}
\label{tab:Combination-example}
\begin{centering}
\begin{tabular}{|c|c|c|}
\hline 
possibility ID & $Z_1$ & $Z_2$ \tabularnewline
\hline  \hline 
1 & $j_1 j_2$ & $j_3 j_4$\tabularnewline \hline 
2 & $j_1 j_2$ & $j_3 j_5$\tabularnewline \hline 
3 & $j_1 j_2$ & $j_4 j_5$\tabularnewline \hline 
4 & $j_1 j_3$ & $j_2 j_4$\tabularnewline \hline 
5 & $j_1 j_3$ & $j_2 j_5$\tabularnewline \hline 
6 & $j_1 j_3$ & $j_4 j_5$\tabularnewline \hline 
... & ... & ...\tabularnewline \hline 
\end{tabular}$\quad\quad$ %
\begin{tabular}{|c|c|c|}
\hline 
possibility ID & $Z_1$ & $Z_2$ \tabularnewline
\hline \hline 
1 & $j_1j_2$ & \sout{$j_3j_4$}\tabularnewline \hline 
2 & $j_1j_2$ & $j_3j_5$\tabularnewline \hline 
3 & $j_1j_2$ & \sout{$j_4j_5$}\tabularnewline \hline 
4 & $j_1j_3$ & $j_2j_4$\tabularnewline \hline 
5 & $j_1j_3$ & $j_2j_5$\tabularnewline \hline 
6 & $j_1j_3$ & $j_4j_5$\tabularnewline \hline 
... & ... & ...\tabularnewline \hline 
\end{tabular}
\par\end{centering}
\end{table}

This combination example can be written in \cl as: 

\begin{verbatim}
object hZs : COMB( jets[-1] jets[-2] ) alias ahz #
the candidate is temporarily called ahz}
    select {ahz}AbsEta < 3.0 
    select {jets[-2]}Pt > 2.1 
    select {jets[-1]}Pt > 5.1 
    select {jets[-1], ahz }dR < 0.6 # dR between ahz and its constituent 1, apply to all 
    select {ahz}m [] 10 200 
\end{verbatim}

In order to activate this new object, and eliminate the combinations that do not satisfy the requirements, one has to put a selection command into the running algorithm (or region); this could be, for example, to have at least two hadronic Z candidates per event:

\begin{verbatim}
algo testCombinations 
    select Size(jets) >= 2 #we need at least 2 jets for a Z boson
    select Count(hZs) >= 2 #the object name is used, not the temporary alias.
\end{verbatim}

As indicated by Table \ref{tab:Combination-example} right side, there are still multiple possibilities, such as rows 2, 4 and 5. To further reduce these by killing the overlapping candidates and leave a single valid one, some sort of ideal condition should be specified. This can be achieved using the previously discussed $\chi^{2}$ minimization. As an example case, let us require the masses of both candidates to be as close as possible to the known Z mass. Now, the final algorithm is given as:

\begin{verbatim}
object hZs : COMB( jets[-1] jets[-2] ) alias ahz #
the candidate is temporarily called ahz
    select { ahz }AbsEta < 3.0 
    select {jets[-2] }Pt > 2.1 
    select {jets[-1] }Pt > 5.1 
    select {jets[-1], ahz }dR < 0.6 # dR between  ahz and its constituent 1, apply to all 
    select { ahz }m [] 10 200 

define zham : {hZs[-1]}m 
define zhbm : {hZs[-2]}m 
define chi2 : (zham - 91.2)^2 + (zhbm - 91.2)^ 2}

region testCombinations 
    select ALL # to count all events # count number size are all the same.
    select Size(jets) >= 2 # we need at least 2 jets for a Z boson 
    select Count(hZs) >= 2 # we need at least 2 Zhad candidates.
    select chi2 ~= 0 \# we kill here overlapping candidates. .

\end{verbatim}

\subsubsection{Looping over a subset of the object collection}
\label{app:loopoversubset}

By default, \cl loops over all objects in a given collection.  However, sometimes it is necessary to loop only over a subset, such as looping only through the first 3 jets.  ADL and \cl allow to specify the subset, e.g. as {\tt jets[0:3]}.

\subsubsection{Minimum and maximum of object attributes }
\label{app:maxmin}

Looping over objects can be used for selecting the minimum or maximum of a function based on any object attribute. An explanatory example could be to apply a selection based on the minimum value of the angular separation between each of the three most energetic jets and the most energetic electron. In \cl, this criterion can be expressed as:
\begin{verbatim}
    select  Min( dR(JET[0:2], ELE[0] )) > 0.9  .
\end{verbatim}

\subsubsection{Summing object attributes }
\label{app:sum}

\cl allows looping over an attribute to calculate the sum of their values. A typical example would be the sum of transverse momenta of a set of jets. Although this frequently used function is predefined and available as HT, it could also be written as:
\begin{verbatim}
    select  Sum( pT(JET) ) >= 20   .
\end{verbatim}

\subsubsection{Object constituents}
\label{app:constituents}

Sometimes, the analysis might necessitate a selection based on jet constituents. \cl allows the modifier word \texttt{constituents} only in case of jets (or any other jet-like objects, such as the large radius FatJets) to refer to these.  An example for defining a new jet object based on criteria on the constituents would be:
\begin{verbatim}
object goodJet using JET
    select q(JET constituents ) == 0  #select neutral constituents
    select Sum(pT(JET constituents ) ) < 40  # PT from remaining constituents
\end{verbatim}
Here the first criterion removes all the charged constituents of each jet and eventually the jet itself if it has no more constituents left, whereas the second criterion imposes an upper limit of 40~GeV to the sum of the transverse momenta of the remaining constituents of each jet. All other functions available in \cl would work in the same way.

\subsubsection{Daughter particles }
\label{app:daughters}

While defining a new particle based on MC truth information, it is sometimes necessary to access the daughters of a given particle.  \cl is capable of accessing the daughters of of an MC truth particle.  
In the following example, the first selection criterion filters the particles that decayed into two or more daughters, while the second criterion is used to select only the daughters with electric charge. 
 
\begin{verbatim}
object DVcandidates take GEN
    select daughters( GEN ) > 1     # 1 child not accepted
    select abs(q(GEN daughters) ) > 0  # charged daugters only
\end{verbatim}

\subsubsection{Hit and miss method}
\label{app:applyHM}

The \texttt{ApplyHM} function can be used to define new objects which pass or fail the efficiency test in that particular region of the parameter space. In \cl, the random number generation is achieved via the TRandom3~\cite{trandom3} function in ROOT libraries. This function reports the time cost of the call to be about 5~ns on an Intel i7 CPU running at 2.6~GHz. 
 
An example for electrons recorded by an imaginary detector whose electron detection efficiency is described by a table called \texttt{myDet} can be written as:
\begin{verbatim}
object myElectron
    take ELE
    select applyHM( myDet({ELE}pT , {ELE}Eta) == 1) # 0 to reject, 1 to accept.
\end{verbatim}
The analysis algorithm can make use of this newly defined object, \texttt{myElectron}, to apply selection criteria, such as the available number of electrons per event etc.

\subsection{Manipulating Events}

\subsubsection{Selecting or rejecting events} 

The conditions based on which an event can be selected or rejected are written in the {\tt region / algo} blocks.  They start with the {\tt select} or {\tt reject} keywords, and are expressed in the form of functions applied on particles complemented by a comparison operator and a limit value. An example for {\tt select} would be 
\begin{verbatim}
    select Size(goodEle) >= 2}  
\end{verbatim}
The synonyms {\tt cut} and {\tt cmd} can be interchangeably used in place of the {\tt select} keyword.  The keyword {\tt reject} is equivalent to {\tt select not}, thus rejecting the events that match the given criteria, as in the example below:
\begin{verbatim}
    reject  {ELE[0] ELE[1]}q == 0 AND {ELE[0] ELE[1]}m [] 80 100
\end{verbatim}
There are also some special keywords that require further discussion. These are
shown in Table \ref{tab:Special-Conditions}. Select \texttt{ALL } accepts all events, for example it can be used for event counting purposes. The next two are scale factors mostly used in ATLAS related analyses. For other input file formats these scale factors are automatically set to unity.

\begin{table}[h]
\caption{Special Conditions in \cl\label{tab:Special-Conditions}}

\centering{}%
\begin{tabular}{|c|c|c|}
\hline 
Keywords & Example & Explanation\tabularnewline
\hline 
\hline 
\texttt{ALL } & \texttt{select ALL} & accept all events\tabularnewline
\hline 
\texttt{LEPsf } & \texttt{cmd LEPsf} & apply leptonic scale factor to MC events\tabularnewline
\hline 
\texttt{bTagSF} & \texttt{cmd bTagSF} & apply b-jet tagging scale factor to MC events\tabularnewline
\hline 
 &  & \tabularnewline
\hline 
\end{tabular}
\end{table}


\subsubsection{Weighing events}
\label{app:weight}

Many analyses require events to be weighted for cross section and luminosity, for trigger efficiencies, or with various scale factors.  \cl has a mechanism for applying constant event weights or event weights from functions, for which examples are shown below:
\begin{verbatim}
    weight trigEff 0.95
    weight ef2Weight myWeight({ELE_0}pT, {ELE_0}Eta)  # weight 2d
\end{verbatim}
The first command sets the weight of the selected events to 0.95, i.e, if the number of selected events is 1000 in the beginning, now it will be counted as 950.
The second command is a slightly more complicated example as it uses a table which defines the event weight according to two parameters: pT and $\eta$. The event weight is thus obtained from that table according to the attributes of the electron with the highest transverse momentum.

\subsubsection{Saving events}
\label{app:save}

In \cl, it is possible to save the currently surviving events at any stage of the running algorithm.
The events are saved into a ROOT~\cite{ROOT}  file using the command \texttt{save} followed by the user-given file name without the .root extension which is automatically added. It is possible to save multiple times in a single algorithm (region) or multiple algorithms. The events in the output file are saved in the native format of \cl, known as the \texttt{lvl0} file. Therefore an example could be:
\begin{verbatim}
    Save   preselects
\end{verbatim}

\subsection{Bins, counts and histograms}

\subsubsection{Bins}
\label{app:bins}

In analyses dealing with multiple bins for signal and/or background regions, \cl provides a simple way for defining the selections for those bins. The binning of the results should happen as the very last stage of a selection by using the keyword {\tt bin}.  Either the variable and the bin boundaries should be explicitly listed, or multiple bins can be assigned to any variable or function using \cl syntax. These two methods are not mutually exclusive and can define overlapping regions. It is to be noted that for the former, one defines two implicit bins: anything below the first value, and anything above the last value are also recorded separately. Results from binning are both printed (depending on the switches in the initialization section of the ADL file) and recorded as a histogram in the output ROOT file. The examples below illustrate the utilization of the bin definition in an analysis algorithm:
\begin{verbatim}
    bin MET 250 300 500 750 1000 # defining multiple bins simultaneously
    bin Size(bjets) == 1  AND  HT [] 500 1000 # defining a single bin
    bin Size(bjets) == 1  AND  HT [] 1000 1500 # defining a single bin
\end{verbatim}

\subsubsection{Counts}
\label{app:counts}

It is possible to register various signal, background or data counts of a region together with their associated errors. The method to achieve this task is to start the ADL file with the definitions of various count formats. Below are two such examples where for each format type, multiple processes with different names can also be defined.

\begin{verbatim}
countsformat results
    process est, "Total estimated BG", stat, syst
    process obs, "Observed data"

countsformat bgests
    process lostlep, "Lost lepton background", stat, syst
    process zinv, "Z --> vv background", stat, syst
    process qcd, "QCD background", stat, syst
\end{verbatim}

A study described in an ADL file might use data counts or a background estimate or all of these for a statistical analysis. Therefore, the appropriate region has to contain the associated event counts and error information using the correct syntax. It should be consistent with the previous definitions starting with keyword {\tt counts}. Here the counts of each process should be separated by a comma, and the errors can be specified either as symmetrical denoted with the \codeword{+-} sign or asymmetrical denoted with separate \codeword{+} and \codeword{-} signs. An example conforming to above definitions is given below. 

\begin{verbatim}
    counts results 230.0 + 16.0 - 10.0 + 10.0 - 12.0  , 224.0
    counts bgests 105.0 +16.0 - 10.0 +-1.0 , 123.0 +-2.0 +-12.0 , 2.3 +-0.5 +-1.4
\end{verbatim}

Once the analysis run is complete, the user finds in the output file a histogram for each of the defined processes with the name defined in the format commands. These histograms can be recalled and used later during the statistical analysis stage.

\subsubsection{Histograms}
\label{app:histo}

\cl allows defining 1D and 2D histograms for any event variable. The syntax for defining histograms follows closely the notation in ROOT. Any histogram should have a name, like \codeword{h1mReco}, and a list of parameters separated by commas. The explanation of the histogram contents should be given in quotation marks, e.g., \codeword{``Z candidate mass (GeV)"}; the number of bins, lower and upper limits as numbers, e.g. \codeword{100, 0, 200}; and finally the quantity to histogram with the ADL notation, e.g.\codeword{ {ELE_0 ELE_1}m}. A similar syntax is also used for the 2D histograms.  The example below show definitions of 1D and 2D histograms:
\begin{verbatim}
region Wtopmass
    select ... 
    select ...
    hmW,"W mass (GeV)", 70, 50, 150, mW
    hmTop,"Top mass (GeV)", 70, 0, 700, mTop
    hmTopmW,"Top and W mass correlation (GeV)", 50, 50, 150, 70, 0, 700, mW, mTop
\end{verbatim}

Apart from the user-defined histograms, \cl by default automatically fills and saves a cutflow efficiency histogram for each analysis region.  In case binning exists, \cl also saves a histogram with bin counts.  

\begin{figure}
\centering
\includegraphics[scale=0.4]{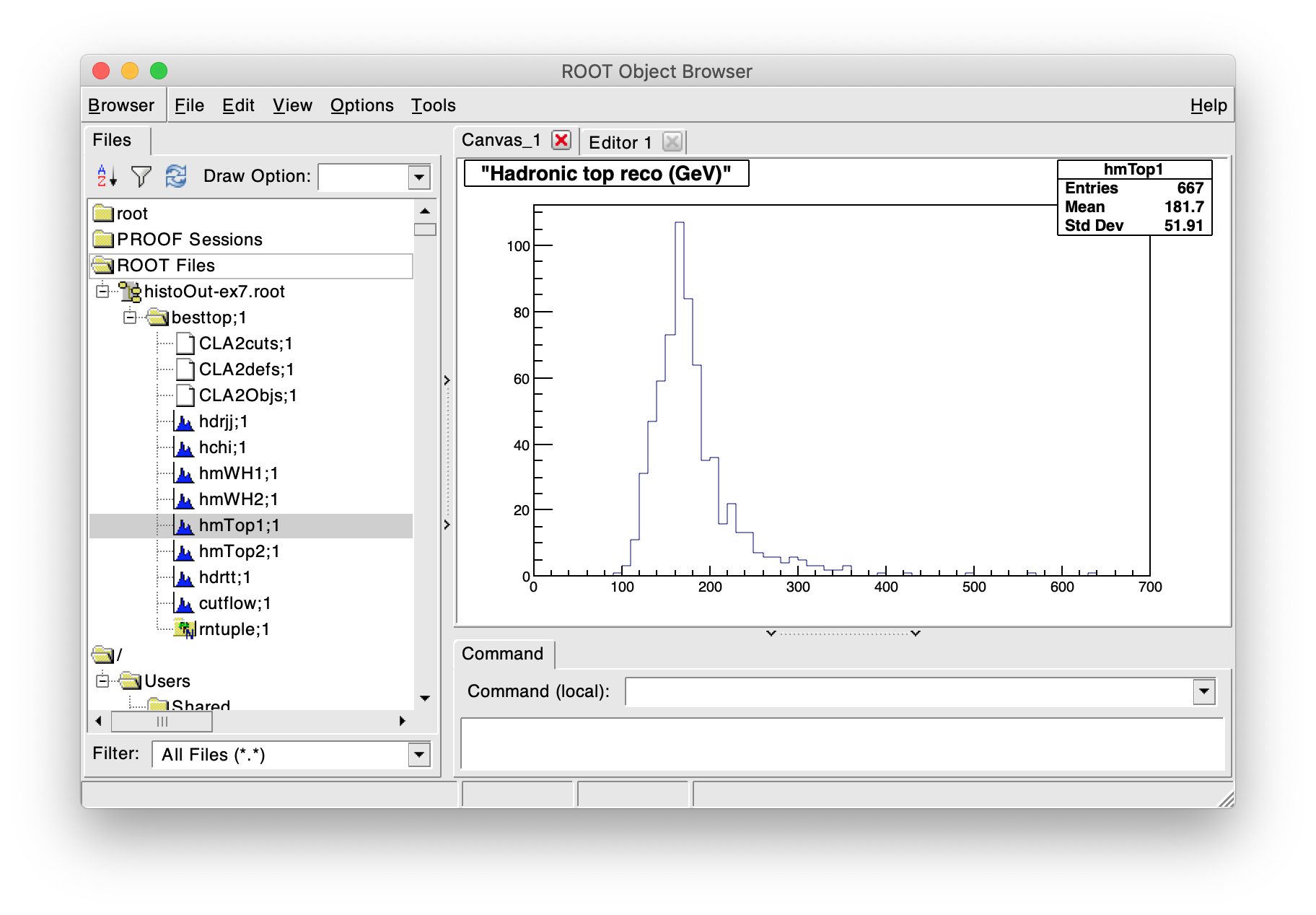}
\caption{An example output from ROOT's TBrowser GUI showing histograms booked and filled by CutLang}
\label{fig:histo-example}
\end{figure}

\blue{Figure~\ref{fig:histo-example} shows a snapshot of the {\tt ROOT TBrowser}, with histograms in an output file listed, and one of the histograms displayed. }

\subsection{Structure of a complete ADL file}

To be run with \cl, an ADL file should follow a definite structure order as described in Section~\ref{sec:adlfilestr}. In this structure, there are mostly optional sections and one compulsory section. The structure order consists of initialization, count format, definitions, new objects, more definitions using new objects, yet newer objects, and event categorization commands. In this list only the event categorization commands are mandatory. The ADL file structure allows multiple concurrent commands to be executed. The details of the first and the last sections are covered next.

\subsubsection{Initialization and information section}
\label{sec:initinfo}
Some of the possible settings in the initialization section have already been discussed in Table \ref{tab:initializations}. It is also possible to include, in this section, some information defining the work that is being done. The allowed keywords and their meaning is explained in the table below.

\begin{table}[h]
\caption{Information keywords of \cl\label{tab:info-section}}

\centering{}%
\begin{tabular}{|r|c|l|}
\hline 
Keywords & Type & Explanation\tabularnewline \hline  \hline 
\texttt{info }       & \texttt{ID} & a name defining the work \tabularnewline \hline 
\texttt{experiment } & \texttt{ID} & a name defining the experiment \tabularnewline \hline 
\texttt{id }         & \texttt{string} & any string defining the work \tabularnewline \hline 
\texttt{title }      & \texttt{string} & any string for the paper title \tabularnewline \hline 
\texttt{publication} & \texttt{string} & any string, the publication information \tabularnewline \hline 
\texttt{sqrtS }      & \texttt{number} & a real number, the collider energy (GeV) \tabularnewline \hline 
\texttt{lumi }       & \texttt{number} & a real number, collected data (fb-1)  \tabularnewline \hline 
\texttt{arXiv }      & \texttt{string} & any string containing the arxiv information  \tabularnewline \hline 
\texttt{hepdata }    & \texttt{string} & any string containing the hepdata information \tabularnewline \hline 
\texttt{doi }        & \texttt{string} & any string containing the doi information \tabularnewline \hline 

\end{tabular}
\end{table}

\subsubsection{Regions and algorithms}

\cl can execute multiple commands in the event categorization section of the ADL file, meaning that the analyst can test multiple methods on the same events independently of each other during design, or work with multiple signal and control regions. The set of commands to be executed for each independent method is called either an algorithm or a region, therefore the keyword to be used is \texttt{algo} or \texttt{algorithm} or \texttt{region} followed with a user selected name, such as:
\begin{verbatim}
    region preselection
\end{verbatim}

Moreover it is possible to define one (1) layer of dependency such that a region can be marked as dependent on another. In this case, the independent region's commands are executed first and the results are saved in a memory cache, and later the dependent region's commands are executed based on that cache. A typical case would be to create multiple signal regions based on a common preselection. This example is illustrated below. Note that the name of the independent region has been used in the dependent region's list of commands directly, without any preceding keywords.

\begin{verbatim}
region preselection
    select ....
    
region SRA
    preselection
    select ....
    
region SRB
    preselection
    select ....
\end{verbatim}

\section{The \cl framework}
\label{appendix:B}
\renewcommand{\thesubsection}{B.\arabic{subsection}}

\subsection{installation and compilation}
The code for the \cl framework can be found in 

\vspace{0.2cm}
\href{https://github.com/unelg/CutLang}{https://github.com/unelg/CutLang}
\vspace{0.2cm}

The ROOT library from CERN should be pre-installed. After downloading the source code, the {\tt make} command should be executed in the {\tt CLA} subdirectory to compiles the whole program. Analyses in \cl are run runs subdirectory using the script {\tt CLA.sh} or {\tt CLA.py}. This subdirectory contains several example files that demonstrate various aspects of ADL and \cl. An analysis can be run using the command
where the input ROOT file  type  can be : {\tt LHCO FCC LVL0 DELPHES ATLASVLL ATLMIN ATLASOD CMSOD CMSNANO }.  The {\tt -i} or {\tt --inifile} option is used for specifying the adl file.

\subsection{External user functions}
\label{app:userfunc}

The addition of the new so called external user functions to the existing set of internal functions is partially automatized. The python helper script {\tt insertExternalFunction.py} in the scripts directory is developed to accomplish this task. It accepts the name of the header file containing the new function as an argument. The automatization currently works with a template based setup, therefore only with certain type of functions. Currently the following input and return types for external functions can be used for building an external function into \cl:
\begin{itemize}
\item receives a vector of TLorentzVectors and an int, returns a vector of TLorentzVector;
\item receives a vector of TLorentzVectors, returns a double;
\item receives a vector of TLorentzVectors and a TVector2, returns a double;
\item receives a vector of TLorentzVectors and a TLorentzVector, returns a double;
\item receives 3 TLorentzVectors, returns a double;
\end{itemize}
The external function must be declared and defined using C/C++ programming language in a header file before running the script.  The script is run with the following command: 
\begin{verbatim}
python insertExternalFunction.py -ext abc
\end{verbatim}
where {\tt abc} is name of the header file without {\tt .h} extension.
Once the helper script runs successfully, the \cl binary has to be recompiled once to use the new function within an ADL file.  

\subsection{Incorporation of new input file types}
\label{app:newformat}

This section describes how to build the interface between a new data file format represented as a 
flat ntuple and the standard types used by the \emph{CutLang} interpreter. This is one aspect of the current version of
\emph{CutLang} that requires some coding expertise.  \emph{CutLang} uses ROOT's \texttt{MakeClass} for this purpose.

\begin{itemize}
\item Obtain a sample ROOT ntuple file containing the new data format and load into ROOT (e.g., using \texttt{TFile f("myfile.root")})
\item Call the ROOT \texttt{MakeClass} command on the relevant tree, specifying a class name
\begin{verbatim}
tree->MakeClass("NewFormatName");
\end{verbatim}
\item Move the resulting header file (\texttt{NewFormatName.h} ) into the
\texttt{analysis\_core} subdirectory, and is include it in the main code \texttt{CLA.Q}  
\item Move the resulting implementation macro (\texttt{NewFormatName.C}) into the
\texttt{CutLang/CLA} directory, and include the following required headers in it:
\begin{lstlisting}
#include <NewFormatName.h>
#include <TH2.h>
#include <TStyle.h>
#include <TCanvas.h>
#include <signal.h>
#include <TObject.h>
#include <TBranchElement.h>

#include "dbx_electron.h"
#include "dbx_muon.h"
#include "dbx_jet.h"
#include "dbx_tau.h"
#include "dbx_a.h"
#include "DBXNtuple.h"
#include "analysis_core.h"
#include "AnalysisController.h"

\end{lstlisting}

\item In the event loop, the input data must be transferred to the standard \cl types, e.g., the electron, muon, photon
and jet particle vectors, without forgetting any available event-wide information like RunNumber, EventNumber etc. An example conversion for the \texttt{LHCO} format is:
\begin{lstlisting} 
TLorentzVector alv; dbxMuon *adbxm; vector<dbxMuon> muons; 
for (unsigned int i=0; i<Muon_; i++) {
  alv.SetPtEtaPhiM(Muon_PT[i], Muon_Eta[i], Muon_Phi[i], (105.658/1E3)); //in GeV
  adbxm= new dbxMuon(alv);
  adbxm->setCharge(Muon_Charge[i] );
  adbxm->setEtCone(Muon_ETiso[i] );
  adbxm->setPtCone(Muon_PTiso[i] );
  adbxm->setParticleIndx(i);
  muons.push_back(*adbxm);
  delete adbxm;
}
\end{lstlisting}

\item Modify the end of the .C file to be as follows:
\begin{lstlisting}
AnalysisObjects a0={muos_map, eles_map, taus_map, gams_map, jets_map, ljets_map, 
              truth_map, combo_map, constits_map, met_map, anevt};

aCtrl.RunTasks(a0);
} // end of event loop
aCtrl.Finalize();
} // end of Loop function
\end{lstlisting}

\item Modify the running script (CLA.sh or CLA.py) to incorporate  the new file format.   
\end{itemize}

\end{document}